\documentclass[10pt,conference, compsocconf]{IEEEtran}

\ifCLASSOPTIONcompsoc
  \usepackage[nocompress]{cite}
\else
  \usepackage{cite}
\fi

\usepackage[pdftex]{graphicx}
\graphicspath{{./}}
\DeclareGraphicsExtensions{.png}

\usepackage{amsmath}
\usepackage{amssymb}
\usepackage{amsthm}
\usepackage{url}

\usepackage{algorithm}
\usepackage{algpseudocode}
\usepackage{verbatim}

\usepackage{lipsum}
\usepackage{multicol}

\usepackage[pdftex]{graphicx}
\usepackage{epstopdf}
\DeclareGraphicsExtensions{.eps}

\usepackage{textcomp}
\usepackage{listings}
\usepackage{color}
\usepackage{mathtools}

\usepackage{interval}
\intervalconfig {
	soft open fences
}

\definecolor{codegreen}{rgb}{0,0.6,0}
\definecolor{codegray}{rgb}{0.5,0.5,0.5}
\definecolor{codepurple}{rgb}{0.58,0,0.82}
\definecolor{backcolour}{rgb}{0.95,0.95,0.92}
 
\lstdefinestyle{mystyle}{
    backgroundcolor=\color{backcolour},   
    commentstyle=\color{codegreen},
    keywordstyle=\color{magenta},
    numberstyle=\tiny\color{codegray},
    stringstyle=\color{codepurple},
    basicstyle=\footnotesize,
    breakatwhitespace=false,         
    breaklines=true,                 
    captionpos=b,                    
    keepspaces=true,                 
    numbers=left,                    
    numbersep=5pt,                  
    showspaces=false,                
    showstringspaces=false,
    showtabs=false,                  
    upquote=true,
    tabsize=2
}
 
\DeclareMathOperator*{\argmin}{arg\,min}

\algnewcommand\algorithmicinput{\textbf{Input:}}
\algnewcommand\INPUT{\item[\algorithmicinput]}
\algnewcommand\algorithmicoutput{\textbf{Output:}}
\algnewcommand\OUTPUT{\item[\algorithmicoutput]}
\algnewcommand{\IIf}[1]{\State\algorithmicif\ #1\ \algorithmicthen}
\algnewcommand{\EndIIf}{\unskip\ \algorithmicend\ \algorithmicif}

\begin{document}

\title{Novel Model-based Methods for Performance Optimization of Multithreaded 2D Discrete Fourier Transform on Multicore Processors}

\author{\IEEEauthorblockN{Semyon Khokhriakov\IEEEauthorrefmark{1},
Ravi Reddy\IEEEauthorrefmark{2},
and
Alexey Lastovetsky\IEEEauthorrefmark{2}}
\IEEEauthorblockA{School of Computer Science\\
University College Dublin\\
Dublin, Ireland\\
Email: \IEEEauthorrefmark{1}semen.khokhriakov@ucdconnect.ie,
\IEEEauthorrefmark{2}\{ravi.manumachu,alexey.lastovetsky\}@ucd.ie}}

\maketitle

\begin{abstract}
Code modernization is a umbrella term used for porting and tuning codes to keep them up-to-date with the rapidly changing hardware landscape and to extract the best performance from the current hardware platforms. Roofline model is generally used to visually depict the cumulative performance gains from optimizations towards achieving the theoretical peak performance of a processor. However, this practice can be retrogressive with two typical symptoms. First, it can fall prey to Red Queen Principle where one spends several man-years putting extensive optimizations only in the long run to stay in the same place where one started. Second, it is very likely that an open source package with portable optimizations may exhibit better average performance overall than a heavily optimized vendor package. 

In this paper, we expound this viewpoint using multithreaded fast Fourier transforms provided in three highly optimized packages, FFTW-2.1.5, FFTW-3.3.7, and Intel MKL FFT. Then, we propose a novel model-based parallel computing technique as a very effective and portable method for optimization of scientific multithreaded routines for performance, especially in the current multicore era where the processors have abundant number of cores. We present two optimization methods, \emph{PFFT-FPM} and \emph{PFFT-FPM-PAD}, based on this technique. They compute 2D-DFT of a complex signal matrix of size $N \times N$ using $p$ abstract processors. Both the algorithms take as inputs, discrete 3D functions of performance against problem size of the processors and output the transformed signal matrix.

Based on our experiments on a modern Intel Haswell multicore server consisting of 36 physical cores, the average and maximum speedups observed for \emph{PFFT-FPM} using \emph{FFTW-3.3.7} are 1.9x and 6.8x respectively and the average and maximum speedups observed using \emph{Intel MKL FFT} are 1.3x and 2x respectively. The average and maximum speedups observed for \emph{PFFT-FPM-PAD} using \emph{FFTW-3.3.7} are 2x and 9.4x respectively and the average and maximum speedups observed using \emph{Intel MKL FFT} are 1.4x and 5.9x respectively.
\end{abstract}


\IEEEpeerreviewmaketitle

\section{Introduction}\label{sec:introduction}

Code modernization is a perpetual endeavour of performance experts to port and tune their codes to keep up-to-date with the rapidly changing hardware platforms and to run efficiently on them. The roofline model is used to visually depict the trend of performance gains accrued from intra-node optimizations towards the theoretical peak performance of a processor. Using this model, the high optimized scientific applications such as Intel Math Kernel Library (Intel MKL) (BLAS, FFT) consistently demonstrate the close-to-peak performance or superior performance of their codes (such as BLAS) for new platforms. However, we show that this practice confined to the specialist domain of code optimization experts can, not only be time-consuming but also harmful in the long run with two typical symptoms. First, it can fall prey to Red Queen Principle where one spends several man-years putting extensive optimizations only in the long run to stay in the same place where one started. This is because hardware architectures are changing rapidly to fuel the progress towards unprecedented computational capabilities such as exascale computing. Architecture-specific optimizations may become obsolete for newer architectures. Second, it is very likely that an open source package with portable optimizations may exhibit better average performance overall than a heavily optimized vendor package. We exemplify this viewpoint using a case study.

We use three multithreaded FFT applications for comparison written using the packages FFTW-2.1.5, FFTW-3.3.7, and Intel MKL FFT respectively. The performance profiles/speed functions for the applications are obtained on a modern Intel Haswell multicore server consisting of 2 sockets of 18 physical cores each (specification shown in Table \ref{table-Haswell-server}). All the FFT applications compute a 2D-DFT of complex signal matrix of size $N \times N$ using 36 threads. We do not use any special environment affinity variables during the execution of the application. The total number of problem sizes $N \times N$ experimented is around 1000 with $N$ ranging from 128 to 64000 with a step size of 64, $\{128, 192,..., 64000\}$. We will be referring frequently to width of performance variations in a performance profile. It is related to the difference of speed between two subsequent local minima ($s_1$) and maxima ($s_2$) and is defined below:

\begin{equation}\label{eq:1}
variation (\%) = \frac{|s_1 - s_2|}{\min(s_1,s_2)} \times 100
\end{equation}

To make sure the experimental results are reliable, we follow a statistical methodology described in the experimental section \ref{experimental-methodology}. Briefly, for every data point in the functions, the automation software executes the application repeatedly until the sample mean lies in the 95\% confidence interval and a precision of 0.025 (2.5\%) has been achieved. For this purpose, Student's t-test is used assuming that the individual observations are independent and their population follows the normal distribution. We verify the validity of these assumptions by plotting the distributions of observations. The speed/performance values shown in the graphical plots throughout this work are the sample means.

Figure \ref{fig:fft215_vs_337_speed}, \ref{fig:fft215_vs_337_avg_speed} show the performance profiles of FFTW 2.1.5 versus FFTW 3.3.7. Following are the key observations:

\begin{figure*}
	\centering
	\includegraphics[width=1\linewidth]{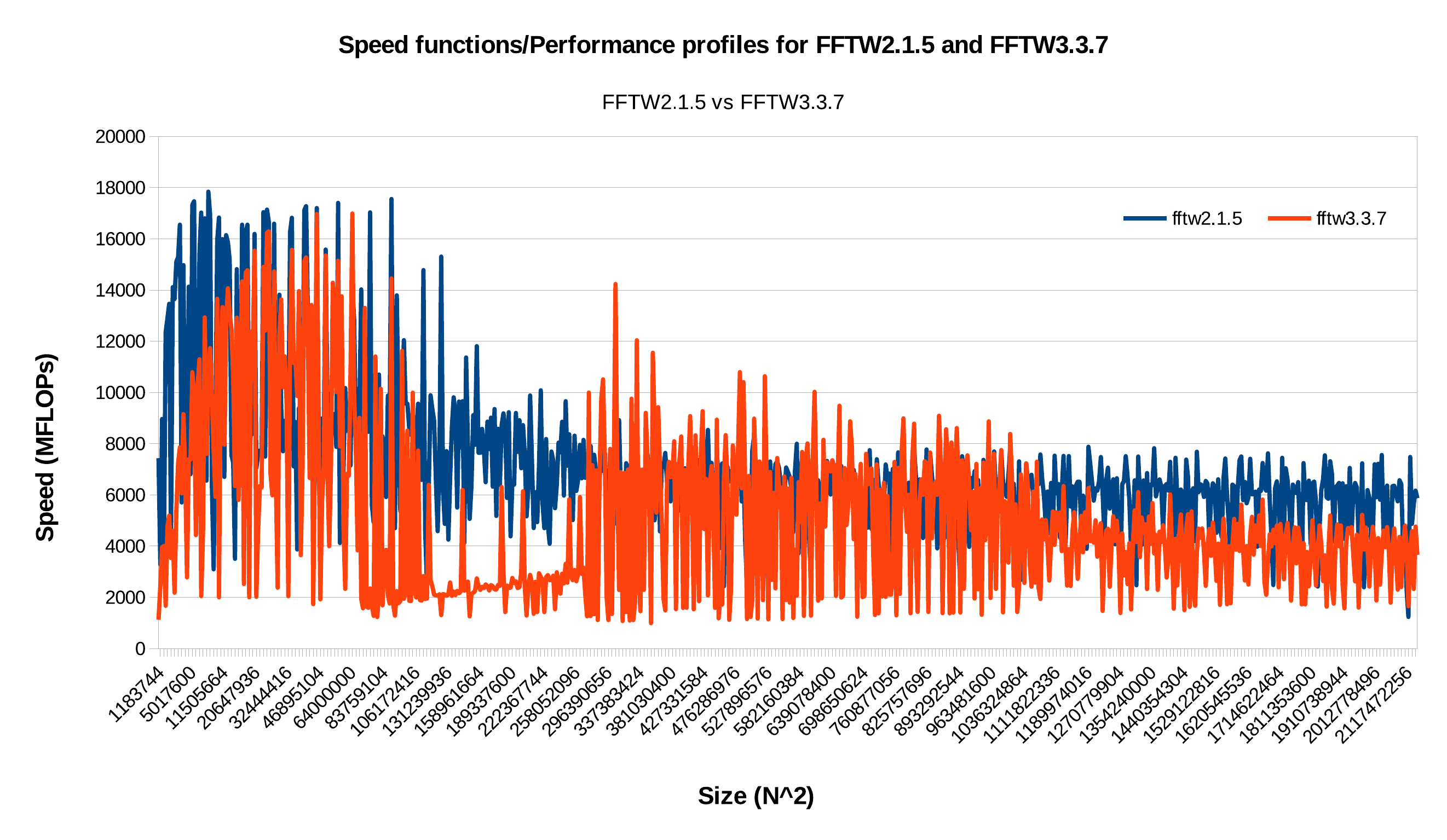}
	\caption[\emph{FFTW-2.1.5 vs FFTW-3.3.7}]{Performance profiles of 2D-FFT computing 2D-DFT of size $N \times N$ using FFTW-2.1.5 and FFTW-3.3.7 respectively. The 2D-FFT applications are executed using 36 threads on a Intel multicore server consisting of two sockets of 18 cores each.}
	\label{fig:fft215_vs_337_speed}
\end{figure*}	
	
\begin{figure*}	
	\centering
	\includegraphics[width=1\linewidth]{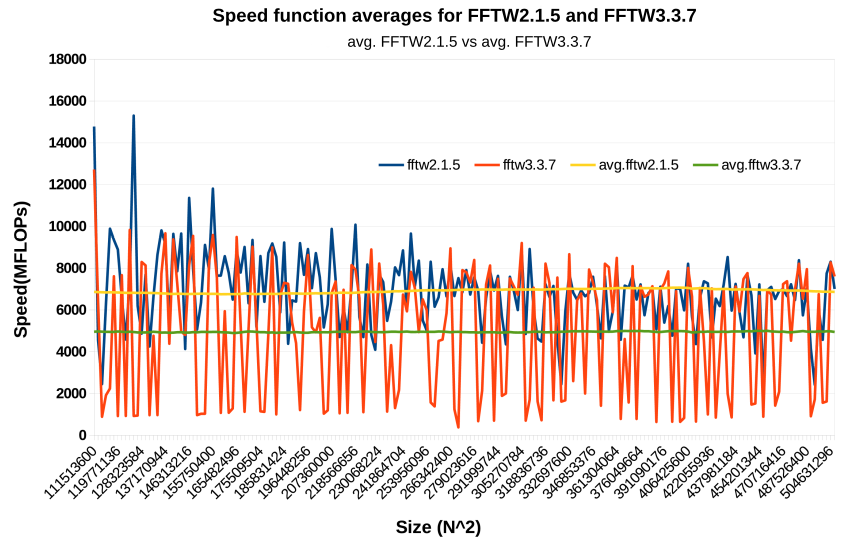}
	\caption[\emph{FFTW-2.1.5 vs FFTW-3.3.7}]{The average speeds of FFTW-2.1.5 vs FFTW-3.3.7 respectively.}
	\label{fig:fft215_vs_337_avg_speed}
\end{figure*}

\begin{itemize}
	  \item We can see that the width of performance variations in FFTW-3.3.7 is substantially greater than that for FFTW-2.1.5. 
	  \item The peak performance of FFTW-3.3.7 is 16989 MFLOPs ($N=8000$) whereas that for FFTW-2.1.5 is 17841 MFLOPs ($N=2816$).
	  \item The average speeds of FFTW-2.1.5 and FFTW-3.3.7 are 7033 MFLOPs and 5065 MFLOPs respectively. FFTW-2.1.5 is better than FFTW-3.3.7 by around 38\% (on an average). There are 529 problem sizes (out of 1000) where the performance of FFTW-2.1.5 is better than FFTW-3.3.7.
\end{itemize}

\begin{figure*}
	\centering
	\includegraphics[width=1\linewidth]{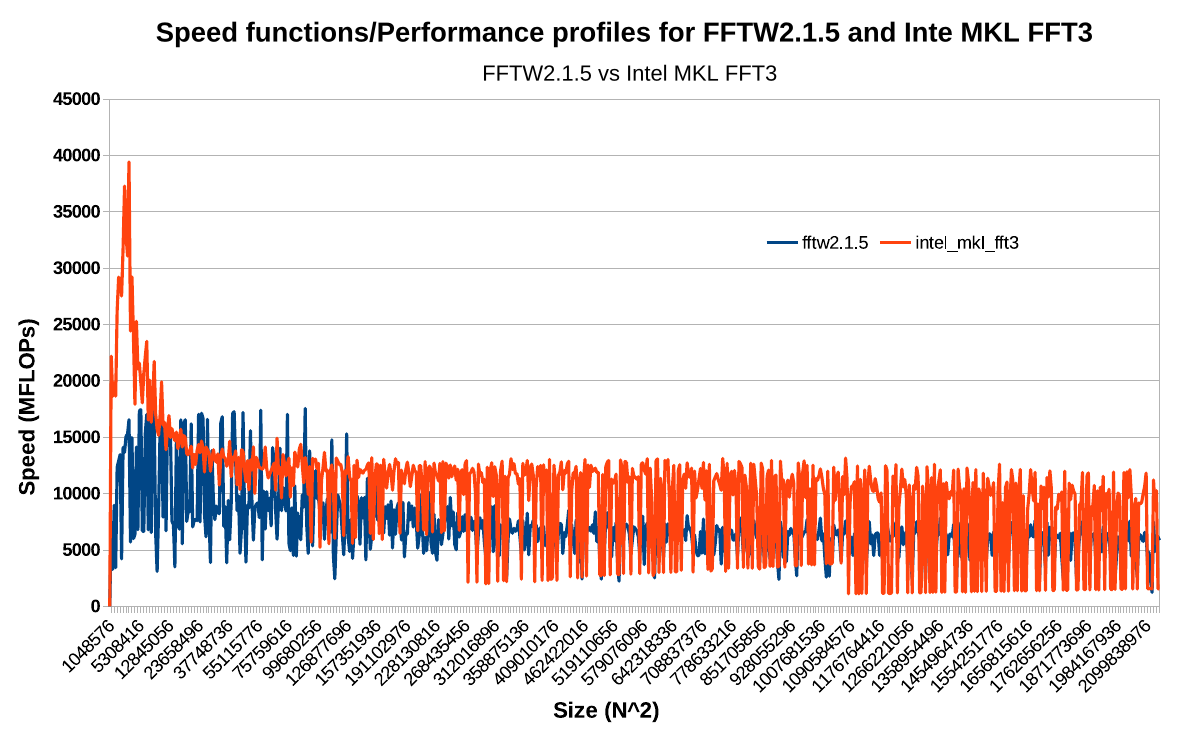}
	\caption[\emph{FFTW-2.1.5 vs Intel MKL FFT}]{Performance profiles of 2D-FFT computing 2D-DFT of size $N \times N$ using FFTW-2.1.5 and Intel MKL FFT respectively. The 2D-FFT applications are executed using 36 threads on a Intel multicore server consisting of two sockets of 18 cores each.}
	\label{fig:fft215_vs_imkl_speed}
\end{figure*}

\begin{figure*}
	\centering
	\includegraphics[width=1\linewidth]{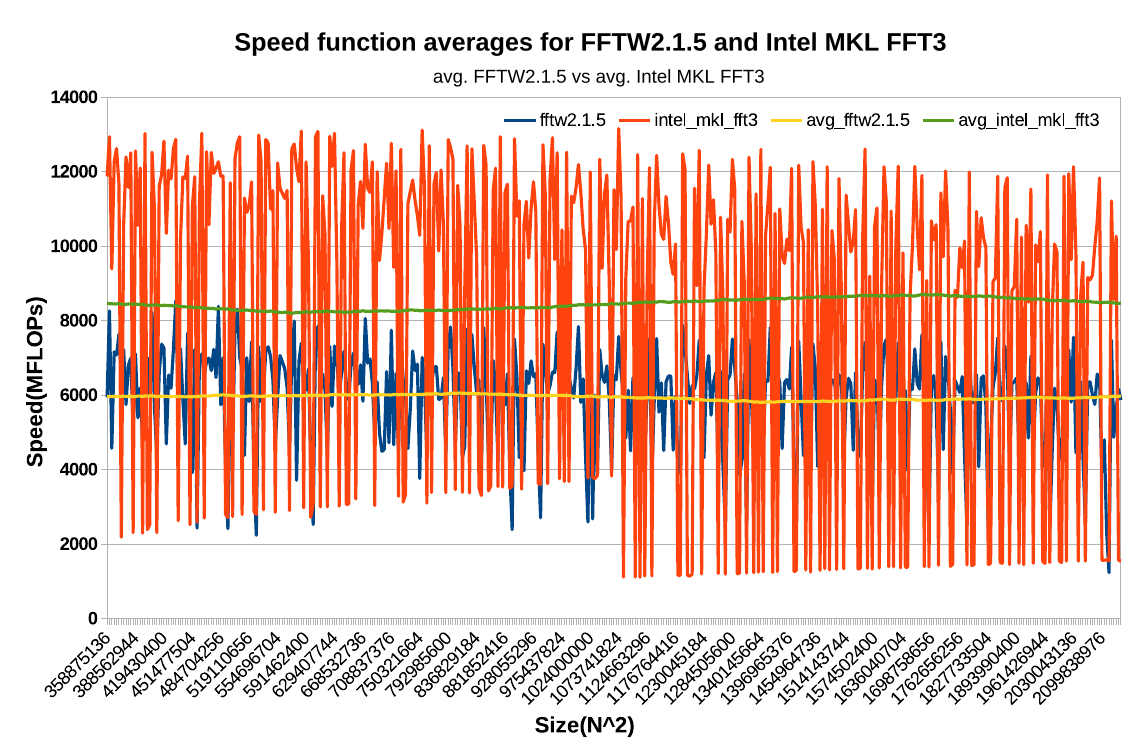}
	\label{fig:fft215_vs_imkl_avg_speed}
	\caption[\emph{FFTW-2.1.5 vs Intel MKL FFT}]{The average speeds of FFTW-2.1.5 and Intel MKL FFT respectively.}
\end{figure*}

Figures \ref{fig:fft215_vs_imkl_speed}, \ref{fig:fft215_vs_imkl_avg_speed} present the performance comparisons between FFTW-2.1.5 and Intel MKL FFT. The most important observations are as follows:

\begin{itemize}
    \item The peak performance of FFTW-2.1.5 is 17841 MFLOPs ($N=2816$) whereas that for Intel MKL FFT is 39424 MFLOPs ($N=1792$).
	\item The average performance of Intel MKL FFT is around 9572 MFLOPs versus 7033 MFLOPs for FFTW-2.1.5. So, on an average, Intel MKL FFT is 36\% better than FFTW-2.1.5. Despite Intel MKL FFT demonstrating better average performance than FFTW-2.1.5, its width of variations is significantly greater than that for FFTW-2.1.5. One can see that the variations of Intel MKL FFT almost fill the picture. This is the reason why Intel MKL FFT demonstrates comparatively poorer average performance despite its high peak performance.
	\item There are 162 problem sizes (out of 1000) where FFTW-2.1.5 is better than Intel MKL FFT.
\end{itemize}

\begin{figure*}
	\centering
	\includegraphics[width=1\linewidth]{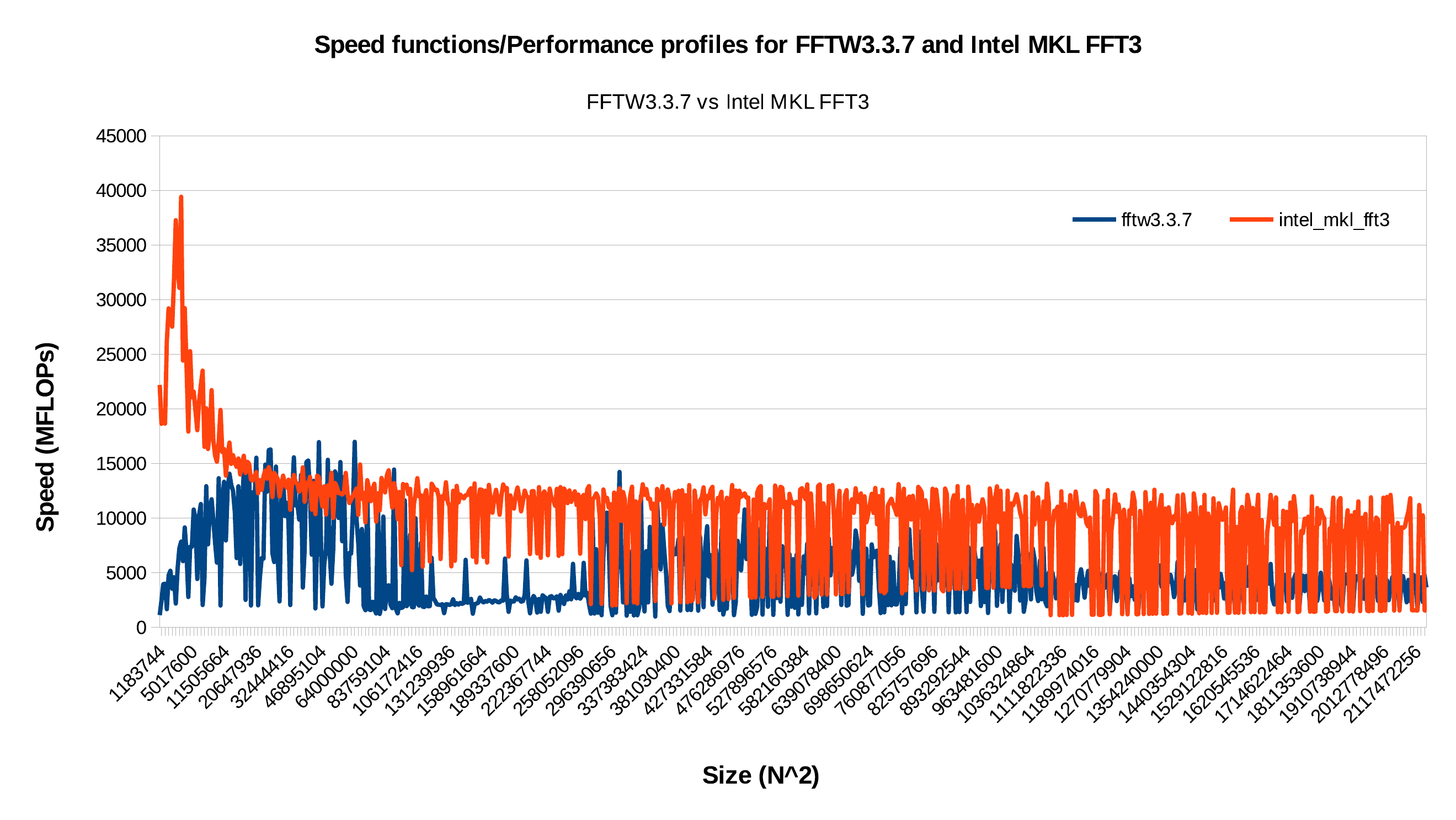}
	\caption[\emph{FFTW-3.3.7 vs Intel MKL FFT}]{Performance profiles of 2D-FFT computing 2D-DFT of size $N \times N$ using FFTW-3.3.7 and Intel MKL FFT respectively. The 2D-FFT applications are executed using 36 threads on a Intel multicore server consisting of two sockets of 18 cores each.}
	\label{fig:fft337_vs_imkl_speed}
\end{figure*}

\begin{figure*}
	\centering	
	\includegraphics[width=1\linewidth]{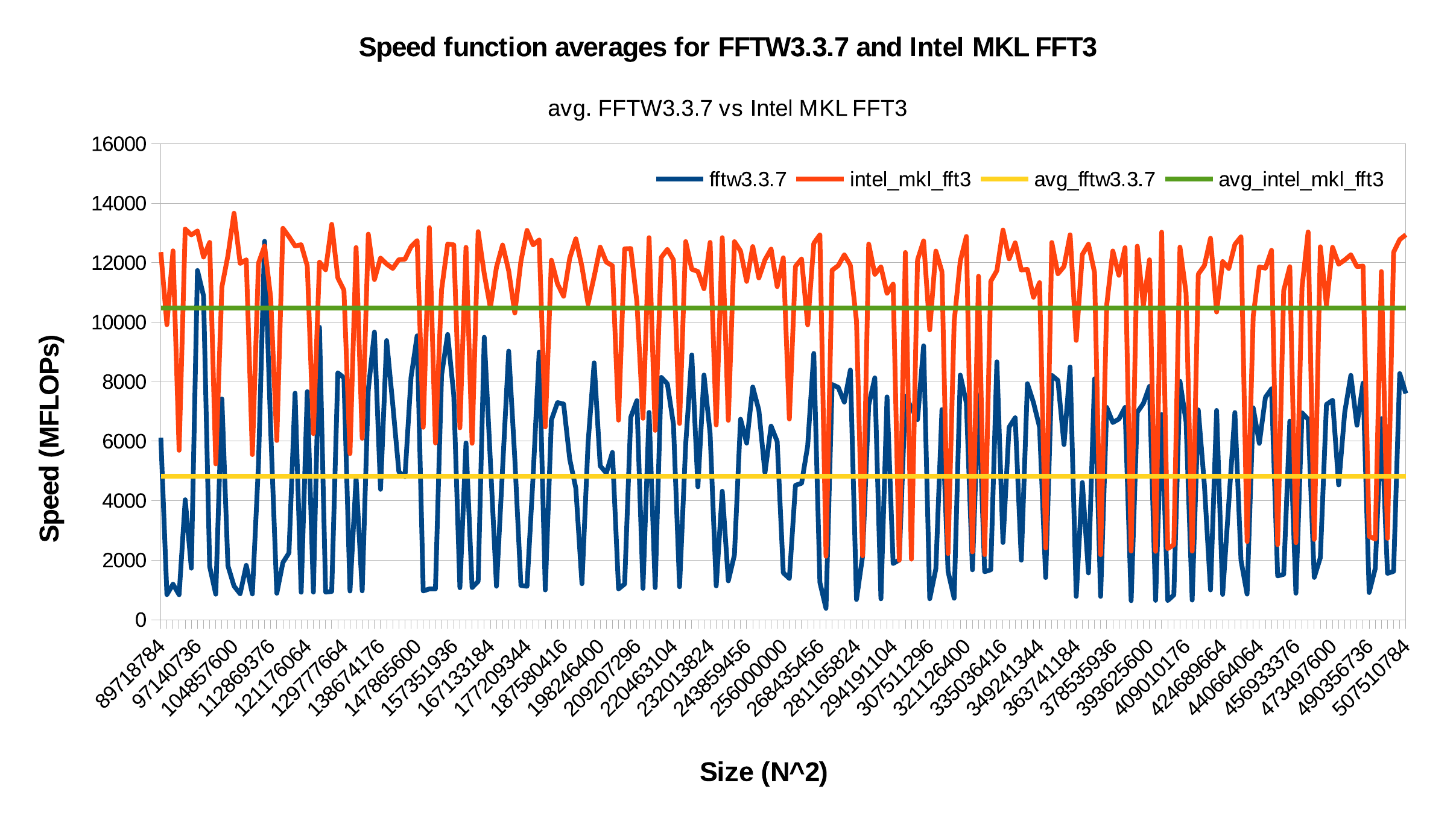}
	\caption[\emph{FFTW-3.3.7 vs Intel MKL FFT}]{The average speeds of FFTW-3.3.7 and Intel MKL FFT respectively.}
	\label{fig:fft337_vs_imkl_avg_speed}
\end{figure*}

Figures \ref{fig:fft337_vs_imkl_speed}, \ref{fig:fft337_vs_imkl_avg_speed} present the performance comparisons between FFTW-3.3.7 and Intel MKL FFT. The crucial observations are as follows:

\begin{itemize}
    \item The peak performance of FFTW-3.3.7 is 16989 MFLOPs ($N=8000$) whereas that for Intel MKL FFT is 39424 MFLOPs ($N=1792$).
	\item The average performance of FFTW-3.3.7 is 5065 MFLOPs and Intel MKL FFT is 9572 MFLOPs, 89\% faster. However, there are 199 problem sizes (out of 1000) where FFTW-3.3.7 outperforms Intel MKL FFT.
	\item The width of variations for Intel MKL FFT is noticeably greater than that for FFTW-3.3.7.
\end{itemize}

Based on these comparisons, we make the following important conclusions:
\begin{itemize}
\item Extensive nodal optimization of FFT using highly architecture-specific techniques is harmful and futile in the long run since hardware platforms undergo drastic changes. A supreme example is FFTW-2.1.5 versus FFTW-3.3.7. The FFT package, FFTW-2.1.5, was last updated in 1999. It outperforms FFTW-3.3.7, which undergoes constant revisions in terms of code optimizations.
\item An open source package may perform better than highly optimized vendor package since it employs portable optimizations. A good example is FFTW-3.3.7 versus Intel MKL FFT. Intel MKL FFT is highly optimized for some specific problem sizes but exhibits poor performance for the rest. This can be seen from the width of its performance variations. Though the average performance of FFTW-3.3.7 is lesser than Intel MKL FFT, it outperforms Intel MKL FFT for many problem sizes and its variations are lesser.
\end{itemize}

There are three solution approaches that can be employed for the optimization of 2D-DFT computation by removal of performance variations. These approaches can be applied, in general, for optimization of data-parallel applications on modern multicore processors for performance. 
\begin{itemize}
\item \textit{Optimization through source code analysis and tuning}: This approach requires source code modification. It lacks portability if architecture-specific optimizations are used. It has other disadvantages, the most crucial being the disproportion between the time spent tuning the code and the continued long-term portable performance improvements.
\item \textit{Optimization using solutions to larger problem sizes with better performance}: This is a portable approach. However, there has to be a performance model, which given workload size $N$ to solve will output the problem size $N_l (> N)$ that is to be used for padding. While programmatically extending 1D arrays logically is easy, it is not the case for 2D arrays such as matrices and multidimensional arrays.
\item \textit{Optimization using model-based parallel computing}: In the current era of multicores where processors have abundant number of cores, one can partition the workload between several identical multithreaded routines (abstract processors) and execute them in parallel. This is a highly portable approach and as we show in this paper, can demonstrate good portable performance.
\end{itemize}
We describe these approaches in the background section \ref{solution-approaches} to follow.

In this paper, we propose a novel model-based parallel computing technique as a very effective and portable method for optimization of multithreaded routines for performance on multicore processors. We present two optimization methods, \emph{PFFT-FPM} and \emph{PFFT-FPM-PAD}, based on this technique. The first method adopts the third solution approach and is a model-based parallel computing solution employing functional performance models (FPMs). The second method is an extension of the first. It combines the second and third approaches where the lengths of the paddings are determined from the FPMs. Both the methods compute 2D-DFT of a complex signal matrix of size $N \times N$ using $p$ abstract processors. They take as inputs, discrete 3D functions of performance against problem size (FPMs) of the processors and output the transformed signal matrix. Unlike load balancing algorithms, optimal solutions found by these algorithms may not load-balance the application. We demonstrate tremendous speedups for both these algorithms over the basic versions offered in FFTW-3.3.7 and Intel MKL FFT.

Our main contributions can be summarized as follows:
\begin{itemize}
    \item We describe the drawbacks inherent in the practice of extensive nodal optimization using highly architecture-specific optimizations. We use computation of 2D-DFT by multi-threaded FFT routines offered by three highly optimized packages, FFTW-2.1.5, FFTW-3.3.7, and Intel MKL FFT, for this purpose. We show that FFTW-2.1.5, which is obsolete and a decade older than FFTW-3.3.7 performs better than it for several problem sizes and has better average performance. We also show that a heavily optimized vendor package, Intel MKL FFT, has severe performance variations compared to FFTW-2.1.5 and FFTW-3.3.7 and several problem sizes where its performance is worse even though its average performance is a bit better.
	\item We propose two novel nodal optimization methods using model-based parallel computing to compute 2D-DFT on modern multicore servers and are therefore highly portable. We report tremendous speedups of these methods over the basic FFT routines provided in the packages FFTW-3.3.7 and Intel MKL FFT. We show that using our optimization methods improves the average performance of FFTW-3.3.7 over the unoptimized FFTW-2.1.5 by 42\% and the average performance of Intel MKL FFT over the unoptimized FFTW-2.1.5 by 24\% (over and above the 36\% of unoptimized Intel MKL FFT).
\end{itemize} 

The rest of the paper is structured as follows. Section 3 presents our two model-based parallel computing solutions. Section 4 contains the experimental results. Section 5 concludes the paper.

\begin{table}
	\centering
	\begin{tabular}{ |c|c| } 
		\hline
		 \textbf{Technical Specifications} & \textbf{Intel Haswell Server} \\ 
		\hline  
		Processor & Intel Xeon CPU E5-2699 v3 @ 2.30GHz \\ 
		\hline
	    OS & CentOS 7.1.1503 \\ 
	    \hline
	    Microarchitecture & Haswell \\ 
	    \hline
	    Memory & 256 GB \\
	    \hline
	    Core(s) per socket & 18 \\	    
	    \hline
	    Socket(s) & 2  \\
	    \hline
	    NUMA node(s) & 2 \\
	    \hline
	    L1d cache & 32 KB \\
	    \hline
	    L1i cache & 32 KB \\
	    \hline
	    L2 cache & 256 KB \\
	    \hline
	    L3 cache & 46080 KB \\
		\hline	
        NUMA node0 CPU(s) & 0-17,36-53 \\
        \hline
        NUMA node1 CPU(s) & 18-35,54-71 \\
		\hline
	\end{tabular}
	\caption{Specification of the Intel Haswell server used to construct the performance profiles.}
	\label{table-Haswell-server}
\end{table}

\section{Performance Optimization of fast Fourier transform on Multicore Processors: Solution Approaches} \label{solution-approaches}

In this section, we describe three solution approaches for the optimization of 2D-DFT computation (by removal of performance variations). These approaches can be applied, in general, for optimization of data-parallel applications on modern multicore processors for performance. We discuss the advantages and disadvantages of each approach.

\textbf{Optimization through source code analysis and tuning}: This is typically the first approach adopted to improve the performance of an application. However, it has many disadvantages.
\begin{itemize}
	\item If the code is highly tuned to a specific vendor architecture, its portability to other vendor architectures suffers. It is also debatable (as we show in this paper) if the performance improvements carry forward to different generations of the same architecture. Therefore, it lacks portable performance.
	\item Most high quality codes are proprietary and therefore their sources are not available for inspection and tuning. For example: BLAS, FFT packages that are part of Intel MKL library.
	\item It will require source code modification. Since the highly optimized packages such as FFTW are written with several man-years of effort for different generations of hardware, any source code change may entail extensive testing to ensure old functionality is not broken. Therefore, it is a time consuming process.
	\item The most crucial disadvantage that we reiterate is the disproportion between the time spent tuning the code and the continued long-term portable performance improvements. Even if the code is available open source, tuning the code requires expertise and intricate knowledge of the hardware architecture. Therefore, code tuning is a highly specialized skill and is also usually time consuming. However, the performance improvements accrued long-term are not always promising. Figures \ref{fig:fft215_vs_337_speed},\ref{fig:fft215_vs_337_avg_speed} depict a striking example showing the performances of FFTW-2.1.5 and FFTW-3.3.7. FFTW-2.1.5 is last updated in 1999 whereas FFTW-3.3.7 is the latest release (September 2017) and contains numerous optimizations (SIMD, AVX, etc.). While for some special problem sizes, FFTW-3.3.7 is better than FFTW-2.1.5, there are many problem sizes where FFTW-2.1.5 outperforms FFTW-3.3.7. The average performance of FFTW-2.1.5 is much better. 
\end{itemize}

\textbf{Optimization using solutions to larger problem sizes with better performance}: Supposing we are solving a problem where the size of the matrix is $N$. In this approach, the solution to a larger problem size ($N_l > N$), which has better execution time than $N$, is used as solution for $N$. The common approach is the pad the input matrix to increase its problem size from $N$ to $N_l$ and zero the contents of the extra padded areas. It is also a technique that is widely used in different flavours (restructuring arrays, aggregation) to minimize cache conflict misses \cite{ishizaka2003cache}, \cite{Zhao2007}, \cite{Hong2016}, \cite{Jiang2017}. It requires no source code modification of the optimized package.

While it is a highly portable approach, it also has some disadvantages.
\begin{itemize}
	\item There has to be a performance model, which given $N$ will provide the problem size $N_l$ that is to be used for padding. In this work, we use functional performance models (FPMs) that will provide this information.
	\item While programmatically extending 1D arrays logically is easy, it is not the case for 2D arrays such as matrices and multidimensional arrays. One inexpensive technique is to locally copy the input signal matrix of size $N$ to a work matrix of size $N_l$, compute 2D-DFT of the work matrix and copy the relevant content back to the signal matrix, which is returned to the user. However, the drawback is the extra memory used for the work matrix.
\end{itemize}

\textbf{Optimization using model-based parallel computing}: Finally, we propose the third approach, which is based on parallel computing. In the current era of multicores where processors have abundant number of cores, one can partition the workload between several identical multithreaded routines (abstract processors) and execute them in parallel. This method can be a very effective nodal optimization technique especially when it employs realistic performance models of computation and efficient data partitioning algorithms that use the models as input.

Its advantages are:
\begin{itemize}
	\item It is highly portable when the performance models of computation used in the data partitioning algorithms do not use architecture-specific parameters.
	\item No source code modification of the optimized package is required.
	\item Relatively less time-consuming programming effort involved, which is to distribute the workload between several identical multithreaded routines (abstract processors) and execute them in parallel.
	\item Speedups can be very good (as we show in this work) and are portable.
\end{itemize}

The disadvantages are:
\begin{itemize}
	\item To distribute the data between the identical multithreaded routines (abstract processors), one can start with homogeneous distribution. But to squeeze out the maximum performance, realistic and accurate performance models and efficient data partitioning algorithms are necessary. It should be noted that the model must not be based on parameters, which are highly architecture-specific (For example: performance monitoring events (PMCs)). This would compromise the portability of this approach.
\end{itemize}

In this paper, we present two algorithms, \emph{PFFT-FPM} and \emph{PFFT-FPM-PAD}. The first algorithm adopts the third approach and is a model-based parallel computing solution employing functional performance models (FPMs). The second is an extension of the first algorithm. It combines the third approach with the second approach where the lengths of the paddings are determined from the FPMs.

\section{2D-DFT: Model-based Parallel Computing Solutions}

In this section, we start with description of the sequential 2D-FFT algorithm using the row-column decomposition method. Next, we explain the parallel 2D-FFT algorithm based on the sequential 2D-FFT algorithm and that uses load balancing technique. Then, we present our two novel model-based optimization methods. The first method \emph{PFFT-FPM} employs parallel computing technique and takes as input, discrete 3D functions of performance against problem size of the processors (FPMs). The second method \emph{PFFT-FPM-PAD} is an extension of \emph{PFFT-FPM} and employs padding, where the partitions (problem sizes) are padded by lengths determined from the FPMs.

\subsection{Sequential 2D-FFT Algorithm}

We first describe the sequential algorithm for computing the DFT on a two-dimensional point discrete signal $\mathcal{M}$ of size $N \times N$. We call $\mathcal{M}$ the signal matrix where each element $\mathcal{M}[i][j]$ is a complex number. The 2D-DFT of $\mathcal{M}$ is defined by:
\begin{align*}
\mathcal{M}[k][l] &= \sum_{i=0}^{N-1}\sum_{j=0}^{N-1} \mathcal{M}[i][j] \times \omega_N^{ki} \times \omega_N^{lj} \\
\omega_N &= e^{-\frac{2\pi}{N}}, 0 \le k,l \le N-1
\end{align*}

The total number of complex multiplications required to compute the 2D-DFT is $\Theta(N^4)$. This complexity can be reduced very significantly by using \textit{row-column decomposition method} where the 2D-DFT is computed using a series of 1D-DFTs, which are implemented using a fast 1D-FFT algorithm. The method consists of two phases called the row-transform phase and column-transform phase. The method is depicted in Figure 4 and is mathematically summarized below:

\begin{align*}
\mathcal{M}[k][l] &= \sum_{i=0}^{N-1}\sum_{j=0}^{N-1} \mathcal{M}[i][j] \times \omega_N^{ki} \times \omega_N^{lj} \\
&= \sum_{i=0}^{N-1} \omega_N^{ki} \times (\sum_{j=0}^{N-1} \mathcal{M}[i][j] \times \omega_N^{lj}) \\
&= \sum_{i=0}^{N-1} \omega_N^{ki} \times (\tilde{\mathcal{M}}[i][l]) \\
&= \sum_{i=0}^{N-1} (\tilde{\mathcal{M}}[i][l]) \times \omega_N^{ki} \\
\omega_N &= e^{-\frac{2\pi}{N}}, 0 \le k,l \le N-1
\end{align*}

It computes a series of ordered 1D-FFTs on the $N$ rows of $x$. That is, each row $i$ (of length $N$) is transformed via a fast 1D-FFT to $\tilde{X}[i][l], \forall l \in [0,N-1]$. The total cost of this row-transform phase is $\Theta(N^2\log_2N)$. Then, it computes a series of ordered 1D-FFTs on the $N$ columns of $\tilde{X}$. The column $l$ of $\tilde{X}$ is transformed to $X[k][l], \forall k \in [0,N-1]$. The total cost of this column-transform phase is $\Theta(N^2\log_2N)$.

Therefore, by using the \textit{row-column decomposition method}, the complexity of 2D-FFT is reduced from $\Theta(N^4)$ to $\Theta(N^2\log_2N)$.

\subsection{\emph{PFFT-LB}: Parallel 2D-FFT Algorithm Using Load Balancing}

\begin{figure*}
	\centering
	\includegraphics[width=\textwidth]{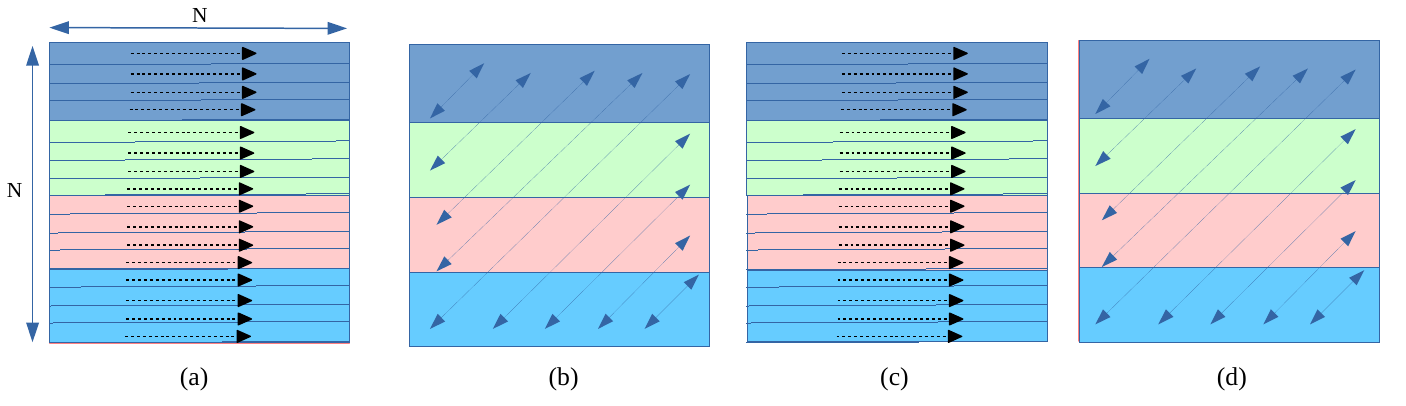}
	\caption{\emph{PFFT-LB} performing 2D-DFT of signal matrix $\mathcal{M}$ of size $N \times N$ ($N=16$) using four identical processors. Each processor is assigned four rows each. (a). Each processor performs series of row 1D-FFTs locally indicated by dotted arrows. (b). Matrix $\mathcal{M}$ is transposed. (a). Each processor performs series of row 1D-FFTs locally indicated by dotted arrows. (d). Matrix $\mathcal{M}$ is transposed again. It is the output of \emph{PFFT-LB}.}
	\label{fig:pfft-lb}
\end{figure*}

The \textit{parallel 2D-FFT algorithm} is based on the sequential 2D-FFT row-column decomposition method and is executed using $p$ identical abstract processors ($\{P_1,...,P_p\}$). The rows of the complex matrix $x$ are partitioned equally between the $p$ processors where each processor gets $\frac{N}{p}$ rows. The other input to the algorithm is the signal matrix $\mathcal{M}$. The output from the algorithm is the transformed signal matrix $\mathcal{M}$. All the FFTs that we discuss in this work are considered to be in-place.

\emph{PFFT-LB} consists of four steps:

\textbf{Step 1. 1D-FFTs on rows:} Processor $P_i$ executes sequential 1D-FFTs on rows $(i-1) \times \frac{N}{p}+1,...,i \times \frac{N}{p}$.
 
\textbf{Step 2. Matrix Transposition:} The matrix $\mathcal{M}$ is transposed.

\textbf{Step 3. 1D-FFTs on rows:} Processor $P_i$ executes sequential 1D-FFTs on rows $(i-1) \times \frac{N}{p}+1,...,i \times \frac{N}{p}$.

\textbf{Step 4. Matrix Transposition:} The matrix $\mathcal{M}$ is again transposed.

The computational complexity of Steps 1 and 3 is $\Theta(\frac{N^2}{p}\log_2N)$. The computational complexity of Steps 2 and 4 is $\Theta(\frac{N^2}{p})$. Therefore, the total computational complexity of \emph{PFFT-LB} is $\Theta(\frac{N^2}{p}\log_2N)$.

The algorithm is illustrated in the Figure \ref{fig:pfft-lb}.

\begin{figure*}
	\centering
	\includegraphics[width=\textwidth]{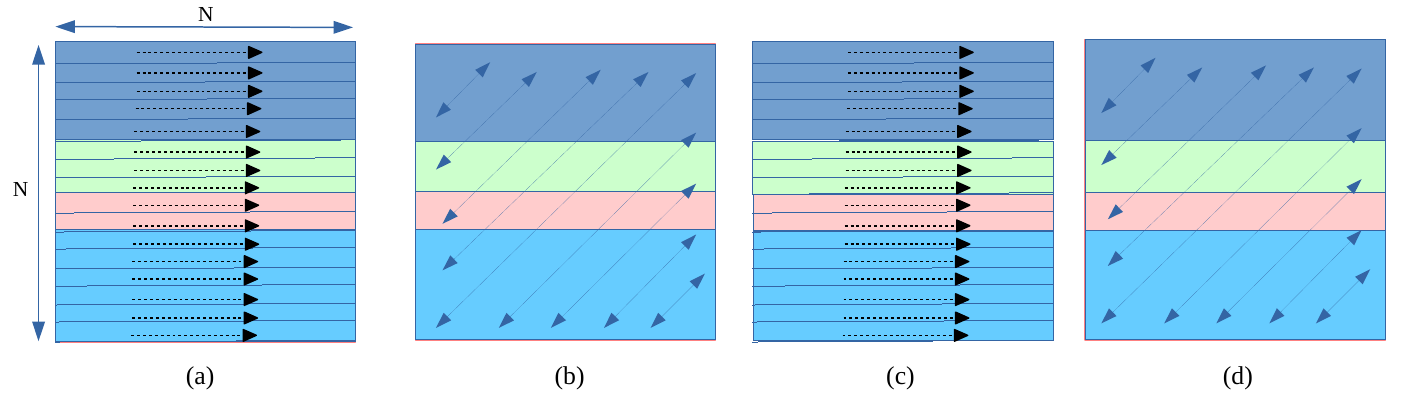}
	\caption{\emph{PFFT-FPM} performing 2D-DFT of signal matrix $\mathcal{M}$ of size $N \times N$ ($N=16$) using four abstract processors. Each processor is assigned different number of rows given by the data distribution, $d=\{5,3,2,6\}$. (a). Each processor performs series of row 1D-FFTs locally indicated by dotted arrows. (b). Matrix $\mathcal{M}$ is transposed. (a). Each processor performs series of row 1D-FFTs locally indicated by dotted arrows. (d). Matrix $\mathcal{M}$ is transposed again. It is the output of \emph{PFFT-FPM}.}
	\label{fig:pfft-fpm}
\end{figure*}

\subsection{\emph{PFFT-FPM}: Performance Optimization Using FPMs and Load Imbalancing}

We now describe our new model-based optimization method called \emph{PFFT-FPM} that employs parallel computing technique and is based on functional performance models (FPMs).

\emph{PFFT-FPM} is executed using $p$ identical abstract processors ($\{P_1,...,P_p\}$). The inputs to \emph{PFFT-FPM} are the number of available abstract processors, $p$, the number of rows of the signal matrix, $N$, the speed functions of the abstract processors, $\mathcal{S}$, and the user-input tolerance $\epsilon$. The output from \emph{PFFT-FPM} is the transformed signal matrix $\mathcal{M}$. The discrete speed function of processor $P_i$ is given by $\mathcal{S}_i=\{s_i(x_1,y_1),...,s_i(x_m,y_m)\})$ where $s_i(x,y)$ represents the speed of execution of $x$ number of 1D-FFTs of length $y$ by the processor $i$. The speed is calculated using the formula: $\frac{2.5 * xy * \log_2(y)}{t}$, where $t$ is the time of execution of $x$ number of 1D-FFTs of length $y$.

It consists of following main steps:

\textbf{Step 1. \textbf{Partition rows:}}

\textbf{\indent 1a. \textbf{Plane intersection of speed functions}:} Speed functions $\mathcal{S}$ are sectioned by the plane $y=N$. A set of $p$ curves on this plane are produced which represent the speed functions against variable $x$ given parameter $y$ is fixed.

\textbf{\indent 1b. \textbf{Are speed functions identical?}:} $\exists (x_k,N), 1 \le k \le m, \quad (\frac{\max_{i=1}^ps_i(x_k,N) - \min_{i=1}^ps_i(x_k,N)}{\min_{i=1}^ps_i(x_k,N)} > \epsilon)$, go to Step 1d. Otherwise, go to Step 1c. If there exists a $(x_k,N)$, the speed functions can not be considered identical.

\textbf{\indent 1c. \textbf{Partition rows using \emph{POPTA}}:} Construct a speed function $S_{avg}=\{s_{avg,i}(x)\},\forall i \in [1,m]$, where $s_{avg,i}(x)=\frac{p}{\sum_{j=1}^{p}\frac{1}{s_j(x,N)}}$. \emph{POPTA} \cite{ravilastov2017} is then invoked using this speed function as an input to obtain an optimal distribution of the rows, $d$.

\textbf{\indent 1d. \textbf{Partition rows using \emph{HPOPTA}}:} \emph{HPOPTA} \cite{hamid2018} is invoked using the $p$ speed curves as input to obtain an optimal distribution of the rows, $d$.

\textbf{Step 2. 1D-FFTs on rows:} Processor $P_i$ executes sequential 1D-FFTs on its rows given by $\{\sum_{k=1}^{i-1}d[i]+1,\cdots,\sum_{k=1}^{i}d[i]\}$.
 
\textbf{Step 3. Matrix Transposition:} The matrix $\mathcal{M}$ is transposed.

\textbf{Step 4. 1D-FFTs on rows:} Same as Step 2.

\textbf{Step 5. Matrix Transposition:} Same as Step 3.

The algorithm is illustrated in the Figure \ref{fig:pfft-fpm} for four abstract processors solving 2D-DFT of size $N \times N (N=16)$. 

The data partitioning algorithms \emph{POPTA} and \emph{HPOPTA} are described in detail in Lastovetsky et al. \cite{ravilastov2017} and Khaleghzadeh et al. \cite{hamid2018} respectively. Briefly, \emph{POPTA} determines the optimal data distribution for minimization of time for the most general performance profiles of data parallel applications executing on homogeneous multicore clusters. One of its inputs is a speed function of the processors involved in its execution since they are considered to be identical. \emph{HPOPTA} is the extension of \emph{POPTA} for heterogeneous clusters of multicore processors. The inputs to it are the $p$ different speed functions of the $p$ processors involved in its execution. Unlike load balancing algorithms, optimal solutions found by both these algorithms may not load-balance an application. The output from the data partitioning algorithms is the data distribution of the rows, $d=\{d_1,\cdots,d_p\}$.

\begin{figure*}
	\centering
	\includegraphics[width=1\linewidth]{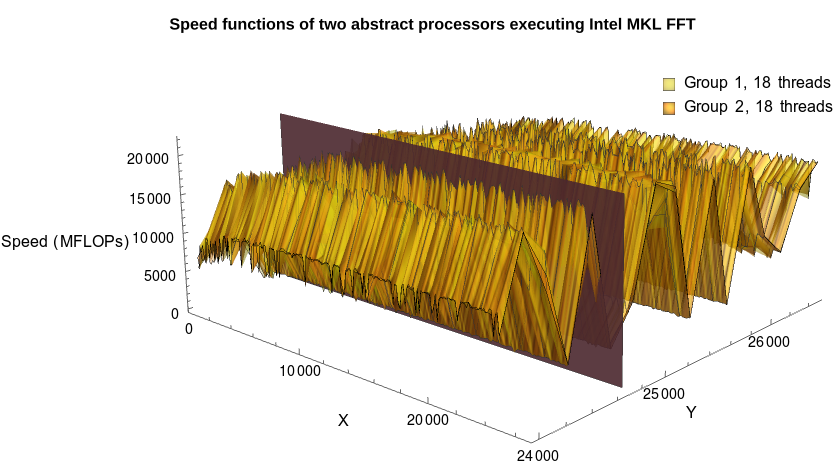}
	\label{fig:fft_intel_mkl_fpms}
	\caption[\emph{PFFT-FPM}]{Speed functions of two abstract processors, each a group of 18 threads. Each group executes 2D-DFT of size $x \times y$ using Intel MKL FFT on a Intel multicore server consisting of two sockets of 18 cores each. Speed functions are intersected by the plane $y=N=24704$.}
\end{figure*}

\begin{figure*}
	\centering
	\includegraphics[width=1\linewidth]{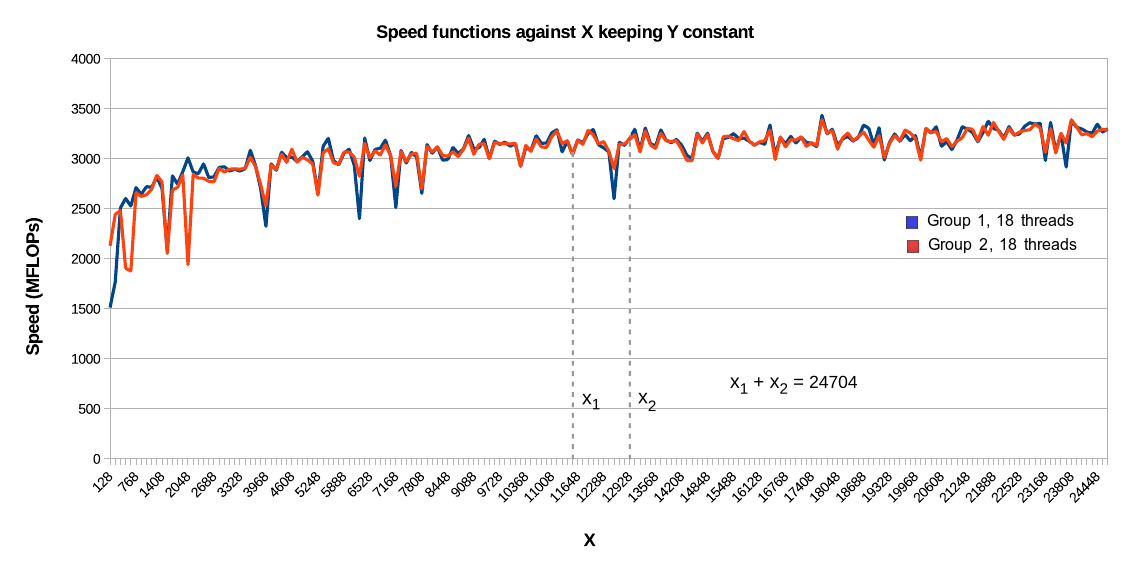}
	\label{fig:fft_intel_mkl_curves}
	\caption[\emph{PFFT-FPM}]{Each intersection produces two curves for the two groups showing speed versus $x$ keeping $y=N=24704$. Application of \emph{HPOPTA} to determine optimal distribution of rows provides the partitioning, $(d[1]=x_1=11648,d[2]=x_2=13056)$.}
\end{figure*}

Figures \ref{fig:fft_intel_mkl_fpms}, \ref{fig:fft_intel_mkl_curves} illustrate the data partitioning algorithm employed in \emph{PFFT-FPM} for two abstract processors solving 2D-DFT of size $N \times N$ where $N=24704$ using Intel MKL FFT on a Intel multicore server. The speed functions shown are segments of the full functions (given in the experimental section \ref{full-speed-functions}). Each abstract processor consists of 18 threads. Figure \ref{fig:fft_intel_mkl_fpms} shows a plane $y=N=24704$ intersecting the two speed functions $\mathcal{S}=\{S_1,S_2\}$ producing two curves, one for each group showing speed versus $x$ given $y=N=24704$. One can see that the two curves are not identical (heterogeneous). That is, there are points where the speeds differ from each other by more than 5\% ($\epsilon = 0.05$). We input the speed functions to \emph{HPOPTA}, which determines the optimal partitioning of rows, $(d[1],d[2])=(11648,13056)$, where each row is of length $N=24704$.

In the following section \ref{pcodes}, we present the pseudocode of \emph{PFFT-FPM} and its shared-memory implementations for Intel MKL FFT and FFTW-3.3.7 respectively on a Intel Haswell server containing 36 physical cores (specification shown in Table \ref{table-Haswell-server}).

\subsection{\emph{PFFT-FPM-PAD}: Performance Optimization Using Padding Determined from FPMs}

In this section, we present \emph{PFFT-FPM-PAD}, an extension of \emph{PFFT-FPM} where the partitions (problem sizes) are padded by lengths determined from the FPMs. The inputs and the outputs of this method are the same as those for \emph{PFFT-FPM}. The data partitioning algorithms invoked in \emph{PFFT-FPM-PAD} are the same as those employed in \emph{PFFT-FPM}. However, the series of 1D-FFTs are performed locally on rows whose length is extended (padded) by an extent determined from the FPM of the processor. It should be noted that the determination of the length of padding is a local computation and is specific to an abstract processor. That is, the lengths can be different for different processors. In some cases, there is no necessity for padding and therefore the length of the padding is zero.

\emph{PFFT-FPM-PAD} consists of following main steps:

\textbf{Step 1. Partition rows:} This step is the same as that for the Algorithm \emph{PFFT-FPM}.

\textbf{Step 2. 1D-FFTs on \textit{padded} rows:} Processor $P_i$ executes sequential 1D-FFTs on its rows in $\mathcal{M}$ given by $d[i]$. The length of each row $N$ is padded to $N_{padded}$. It is determined as follows using the FPM, $\mathcal{S}_i=s_i(x,y)$:
\begin{equation*}
N_{padded} = \argmin_{\mathcal{V} \in \interval[open left]{y_N}{y_m}} (\frac{d[i] \times \mathcal{V}}{s_i(d[i],\mathcal{V})} < \frac{d[i] \times N}{s_i(d[i],N)})
\end{equation*}
The ratio $\frac{x \times y}{s_i(x,y)}$ gives the execution time of problem size $x \times y$. Essentially we select the point in the range \{$(d[i],y_{N+1}),...,(d[i],y_m)$\} that has minimal execution time and better execution time than the point $(d[i],N)$. If no such point is found, the padding length is set to 0. The elements in the padded region $\mathcal{M}[*,c],\forall c \in [N+1,\mathcal{V}]$ are set to 0.

\textbf{Step 3. Matrix Transposition:} The matrix $\mathcal{M}$ (excluding the padded region) is transposed.

\textbf{Step 4. 1D-FFTs on \textit{padded} rows:} The lengths of the paddings already determined in Step 2 are reused. Processor $P_i$ executes sequential 1D-FFTs on its padded rows.

\textbf{Step 5. Matrix Transposition:} Same as Step 3.

\begin{figure*}
	\centering
	\includegraphics[width=1\linewidth]{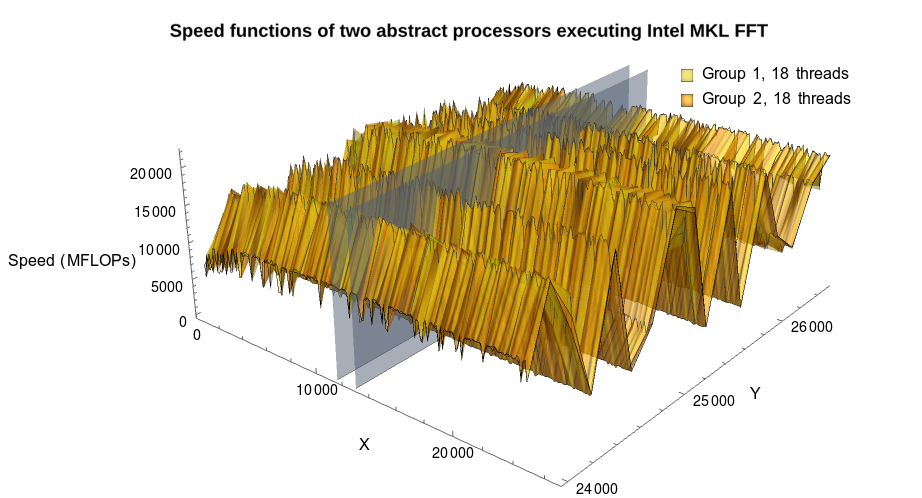}
	\label{fig:fft_intel_mkl_fpms_pad}
	\caption[\emph{PFFT-FPM-PAD}]{Speed function for group1 intersected by the plane $x_1=11648$. Speed function for group2 intersected by the plane $x_2=13056$.}	
\end{figure*}

\begin{figure*}
	\centering
	\includegraphics[width=1\linewidth]{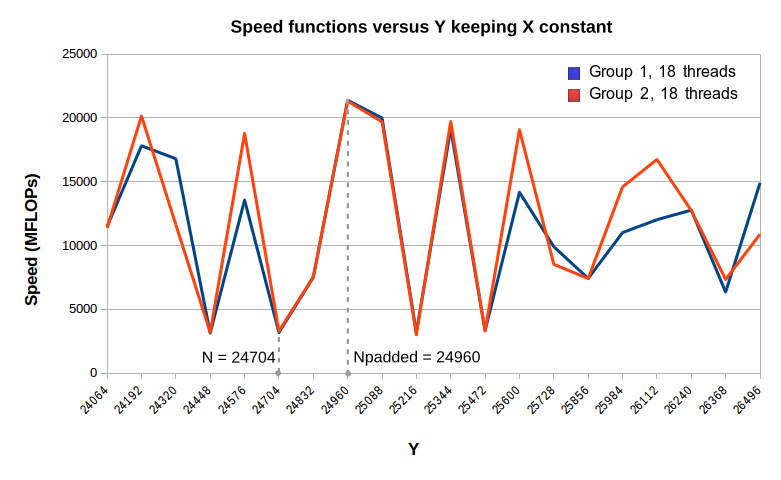}
	\label{fig:fft_intel_mkl_pad_curves}
	\caption[\emph{PFFT-FPM-PAD}]{Each intersection produces a curve for the group showing speed versus $y$ keeping $x$ constant. The lengths of padding for the two groups, $Npadded$, is the same and is equal to 24960.}
\end{figure*}

All the steps of \emph{PFFT-FPM-PAD} are the same as \emph{PFFT-FPM} except the determination of the lengths of the paddings. Figures \ref{fig:fft_intel_mkl_fpms_pad}, \ref{fig:fft_intel_mkl_pad_curves} illustrate how they are determined from the FPMs for two abstract processors solving 2D-DFT of size $N \times N$ where $N=24704$ using Intel MKL FFT on a Intel multicore server. The speed functions shown are segments of the full functions (given in the experimental section \ref{full-speed-functions}). Each abstract processor consists of 18 threads. Figure \ref{fig:fft_intel_mkl_fpms_pad} shows two planes $x_1=11648$ and $x_2=13056$ intersecting the two speed functions $\mathcal{S}=\{S_1,S_2\}$ producing two curves, one for each group showing speed versus $y$ keeping $x$ constant. The padded lengths ($N_{padded,1},N_{padded,2}$) corresponding to $x_1$ and $x_2$ are determined from the curves and are equal to 24960.

In the following section \ref{pcodes}, we present the pseudocode of \emph{PFFT-FPM-PAD} and its shared-memory implementations for Intel MKL FFT and FFTW-3.3.7 respectively on the Intel Haswell server containing 36 physical cores (specification is shown in Table \ref{table-Haswell-server}).

\section{Pseudocodes of \emph{PFFT-FPM} and \emph{PFFT-FPM-PAD}} \label{pcodes}

In this section, we describe two shared memory implementations of \emph{PFFT-FPM}, one using \emph{Intel MKL FFT} and the other using \emph{FFTW-3.3.7}. 

The inputs to the implementation are the signal matrix $\mathcal{M}$ of size $N \times N$, the number of abstract processors (groups) $p$, the speed functions represented by a set $\mathcal{S}$ respectively containing problem sizes and speeds, and number of threads in each abstract processor (group) represented by $t$. The output from the algorithm is the transformed signal matrix $\mathcal{M}$ (considering that we are performing in-place FFT).

The pseudocode of the algorithm is illustrated in (Algorithm \ref{pfft-fpm}). The first step (Line 1) is to determine the partitioning of rows by invoking the routine \emph{PARTITION}. The partitioning routine checks if the variation of the speeds for each data point is less than or equal to user-input tolerance $\epsilon$ (Algorithm \ref{pfft-partition}, Line 3). If a point exists for which the variation exceeds $\epsilon$, then the data partitioning algorithm \emph{HPOPTA} \cite{hamid2018} is invoked (Line 5) to determine the data partitioning of the rows. If all the variations are less than or equal to $\epsilon$, the average of the speeds are calculated for each data point (Line 7). The averaged speed function is then input to \emph{POPTA} \cite{ravilastov2017} to determine the data partitioning of the rows (Line 9). The data distribution is output in the array, $d=\{d_1,\cdots,d_p\}$.

\begin{algorithm}
	\caption{Parallel algorithm computing 2D-DFT of signal matrix $\mathcal{M}$ of size $N \times N$ employing functional performance models (FPMs).} \label{pfft-fpm}
	\begin{algorithmic}[1]
		\Procedure{PFFT-FPM}{$N,\mathcal{M},p,\mathcal{S},t$}
		\INPUT
		\Statex $\mathcal{M}$, Signal matrix of size $N \times N, N \in \mathbb Z_{> 0}$
		\Statex Number of abstract processors, $p \in \mathbb Z_{> 0}$
		\Statex Functional performance model (speed functions) represented by,
		\Statex $\mathcal{S} = \{S_1,...,S_p\}$,
		\Statex $\mathcal{S}_i = \{(x_i[q][r],s_i[q][r])~|~i \in [1,p], q,r \in [1,m], x_i[q][r] \in \mathbb Z_{> 0}, s_i[q][r] \in \mathbb R_{> 0}\}$
		\Statex User tolerance, $\epsilon \in \mathbb R_{> 0}$
		\OUTPUT
		\Statex $\mathcal{M}$, Signal matrix of size $N \times N, N \in \mathbb Z_{> 0}$
		\Statex
		\State $d \gets \Call{Partition}{N,p,\mathcal{S},\epsilon,d}$
		\State $\Call{pfft\_limb}{p,d,N,\mathcal{M}}$
		\State \textbf{return} $\mathcal{M}$
		\EndProcedure
	\end{algorithmic}
\end{algorithm}

\begin{algorithm}
	\caption{Data partitioning of rows of signal matrix $\mathcal{M}$ of size $N \times N$ using the FPMs.} \label{pfft-partition}
	\begin{algorithmic}[1]
		\Procedure{Partition}{$N,p,\mathcal{S},\epsilon,d$}
		\INPUT
		\Statex $N$, Number of rows in the signal matrix, $N \in \mathbb Z_{> 0}$
		\Statex Number of abstract processors, $p \in \mathbb Z_{> 0}$
		\Statex Functional performance model (speed functions) represented by,
		\Statex $\mathcal{S} = \{S_1,...,S_p\}$,
		\Statex $\mathcal{S}_i = \{(x_i[q][r],s_i[q][r])~|~i \in [1,p], q,r \in [1,m], x_i[q][r] \in \mathbb Z_{> 0}, s_i[q][r] \in \mathbb R_{> 0}\}$
		\Statex User tolerance, $\epsilon \in \mathbb R_{> 0}$
		\OUTPUT
		\Statex Optimal partitioning of the rows of the signal matrix, $d=\{d_1,...,d_p\}, d_i \in \mathbb Z_{> 0}, \forall i \in [1,p]$
		\Statex
		\For{$point \gets 1,m$}
		\State $rdiff \gets \frac{\max_{i=1}^ps_i[point][N] - \min_{i=1}^ps_i[point][N]}{\min_{i=1}^ps_i[point][N]}$
		\If{($rdiff > \epsilon$)} 
		\State \textbf{return} $\Call{HPOPTA}{N,p,S,d}$ 
		\EndIf
		\State $S_{avg}[point] \gets \frac{p}{\sum_{i=1}^{p}\frac{1}{s_i[point][N]}}$
		\EndFor
		\State \textbf{return} $\Call{POPTA}{N,p,S_{avg},d}$
		\EndProcedure
	\end{algorithmic}
\end{algorithm}

\begin{algorithm}
	\caption{Parallel algorithm computing 2D-DFT of signal matrix $\mathcal{M}$ of size $N \times N$.} \label{pfft-limb}
	\begin{algorithmic}[1]
		\Procedure{PFFT\_LIMB}{$p,d,N,\mathcal{M}$}
		\INPUT
		\Statex $\mathcal{M}$, Signal matrix of size $N \times N, N \in \mathbb Z_{> 0}$
		\Statex Number of abstract processors, $p \in \mathbb Z_{> 0}$
		\OUTPUT
		\Statex $\mathcal{M}$, Signal matrix of size $N \times N, N \in \mathbb Z_{> 0}$
		\Statex
		\For{$proc \gets 1,p$}
		\State $\Call{1D\_ROW\_FFTS\_LOCAL}{proc,d_{proc},N,\mathcal{M}}$
		\EndFor
		\State $\Call{Parallel\_Tranpose}{\mathcal{M}}$
		\For{$proc \gets 1,p$}
		\State $\Call{1D\_ROW\_FFTS\_LOCAL}{proc,d_{proc},N,\mathcal{M}}$        
		\EndFor		
		\State $\Call{Parallel\_Tranpose}{\mathcal{M}}$
		\State \textbf{return} $\mathcal{M}$
		\EndProcedure
	\end{algorithmic}
\end{algorithm}

\begin{algorithm}
	\caption{Intel MKL implementation of PFFT\_LIMB using FFTW interface employing two groups ($p=2$) of $t$ threads each.} \label{pfft_imkl}
	\begin{algorithmic}[1]
		\Procedure{PFFT\_LIMB\_INTEL\_MKL}{$id,d,N,\mathcal{M}$}
		\INPUT
		\Statex $\mathcal{M}$, Signal matrix of size $N \times N, N \in \mathbb Z_{> 0}$
		\Statex Workload distribution, $d = \{d_1,d_2\}, d_1,d_2 \in \mathbb Z_{> 0}$
		\OUTPUT
		\Statex $\mathcal{M}$, Signal matrix of size $N \times N, N \in \mathbb Z_{> 0}$
		\Statex
		\State $\Call{fftw\_init\_threads()}{}$
		\State $\Call{fftw\_plan\_with\_nthreads}{t}$
		\State $\textbf{\#pragma omp parallel sections num\_threads(2)}$
		\State \hspace{\algorithmicindent} $\textbf{\#pragma omp section}$
		\State \hspace{\algorithmicindent}\hspace{\algorithmicindent} $\Call{1d\_row\_ffts\_local}{1,d_1,N,\mathcal{M}}$
		\State \hspace{\algorithmicindent} $\textbf{\#pragma omp section}$
		\State \hspace{\algorithmicindent}\hspace{\algorithmicindent} $\Call{1d\_row\_ffts\_local}{2,d_2,N,\mathcal{M}}$            
		\State $\Call{Tranpose}{\mathcal{M}}$           
		\State $\textbf{\#pragma omp parallel sections num\_threads(2)}$
		\State \hspace{\algorithmicindent} $\textbf{\#pragma omp section}$
		\State \hspace{\algorithmicindent}\hspace{\algorithmicindent} $\Call{1d\_row\_ffts\_local}{1,d_1,N,\mathcal{M}}$
		\State \hspace{\algorithmicindent} $\textbf{\#pragma omp section}$
		\State \hspace{\algorithmicindent}\hspace{\algorithmicindent} $\Call{1d\_row\_ffts\_local}{2,d_2,N,\mathcal{M}}$            
		\State $\Call{Tranpose}{\mathcal{M}}$                          
		\State $\Call{fftw\_cleanup\_threads()}{}$    
		\State \textbf{return} $\mathcal{M}$
		\EndProcedure
	\end{algorithmic}
\end{algorithm}

\begin{algorithm}
	\caption{FFTW implementation of PFFT\_LIMB employing two groups ($p=4$) of $t$ threads each.} \label{pfft_fftw}
	\begin{algorithmic}[1]
		\Procedure{PFFT\_LIMB\_FFTW}{$d,N,\mathcal{M}$}
		\INPUT
		\Statex $\mathcal{M}$, Signal matrix of size $N \times N, N \in \mathbb Z_{> 0}$
		\Statex Workload distribution, $d = \{d_1,d_2,d_3,d_4\}, d_i \in \mathbb Z_{> 0}, \forall i \in [1,4]$
		\OUTPUT
		\Statex $\mathcal{M}$, Signal matrix of size $N \times N, N \in \mathbb Z_{> 0}$
		\Statex
		\State $\Call{fftw\_init\_threads()}{}$
		\State $\Call{fftw\_plan\_with\_nthreads}{t}$
		\State $\textbf{\#pragma omp parallel sections num\_threads(4)}$
		\State \hspace{\algorithmicindent} $\textbf{\#pragma omp section}$
		\State \hspace{\algorithmicindent}\hspace{\algorithmicindent} $\Call{1d\_row\_ffts\_local}{1,d_1,N,\mathcal{M}}$
		\State \hspace{\algorithmicindent} $\textbf{\#pragma omp section}$
		\State \hspace{\algorithmicindent}\hspace{\algorithmicindent} $\Call{1d\_row\_ffts\_local}{2,d_2,N,\mathcal{M}}$            
		\State \hspace{\algorithmicindent} $\textbf{\#pragma omp section}$
		\State \hspace{\algorithmicindent}\hspace{\algorithmicindent} $\Call{1d\_row\_ffts\_local}{3,d_3,N,\mathcal{M}}$
		\State \hspace{\algorithmicindent} $\textbf{\#pragma omp section}$
		\State \hspace{\algorithmicindent}\hspace{\algorithmicindent} $\Call{1d\_row\_ffts\_local}{4,d_4,N,\mathcal{M}}$                       
		\State $\Call{Tranpose}{\mathcal{M}}$
		\State $\textbf{\#pragma omp parallel sections num\_threads(4)}$
		\State \hspace{\algorithmicindent} $\textbf{\#pragma omp section}$
		\State \hspace{\algorithmicindent}\hspace{\algorithmicindent} $\Call{1d\_row\_ffts\_local}{1,d_1,N,\mathcal{M}}$
		\State \hspace{\algorithmicindent} $\textbf{\#pragma omp section}$
		\State \hspace{\algorithmicindent}\hspace{\algorithmicindent} $\Call{1d\_row\_ffts\_local}{2,d_2,N,\mathcal{M}}$            
		\State \hspace{\algorithmicindent} $\textbf{\#pragma omp section}$
		\State \hspace{\algorithmicindent}\hspace{\algorithmicindent} $\Call{1d\_row\_ffts\_local}{3,d_3,N,\mathcal{M}}$
		\State \hspace{\algorithmicindent} $\textbf{\#pragma omp section}$
		\State \hspace{\algorithmicindent}\hspace{\algorithmicindent} $\Call{1d\_row\_ffts\_local}{4,d_4,N,\mathcal{M}}$                       
		\State $\Call{Tranpose}{\mathcal{M}}$     
		\State $\Call{fftw\_cleanup\_threads()}{}$              
		\State \textbf{return} $\mathcal{M}$
		\EndProcedure
	\end{algorithmic}
\end{algorithm}

\begin{algorithm}
	\caption{Series of $x$ row 1D-FFTs using FFTW interface function fftw\_plan\_many\_dft.} \label{fft_many_1d}
	\begin{algorithmic}[1]
		\Procedure{1D\_ROW\_FFTS\_LOCAL}{$id,x,N,\mathcal{M}$}
		\INPUT
		\Statex Processor identifier, $id \in \mathbb Z_{> 0}$
		\Statex Problem size $x \in \mathbb Z_{> 0}$
		\Statex $\mathcal{M}$, Signal matrix of size $N \times N, N \in \mathbb Z_{> 0}$
		\OUTPUT
		\Statex $\mathcal{M}$, Signal matrix of size $N \times N, N \in \mathbb Z_{> 0}$
		\Statex
		\State $rank \gets 1$; $howmany \gets x$; $s \gets {N}$;
		\State $idist \gets N$; $odist \gets N$; $istride \gets 1$;
		\State $ostride \gets 1$; $inembed \gets s$; $onembed \gets s$;
		\State $plan \gets \Call{fftw\_plan\_many\_dft}{rank, s, howmany,$ \par
			\hskip\algorithmicindent $\mathcal{M},inembed,istride,idist,$ \par
			\hskip\algorithmicindent $\mathcal{M},onembed,ostride,odist,$ \par
			\hskip\algorithmicindent $FFTW\_FORWARD,FFTW\_ESTIMATE$}
		\State $\Call{fftw\_execute}{plan}$
		\State $\Call{fftw\_destroy\_plan}{plan}$    
		\State \textbf{return} $\mathcal{M}$    
		\EndProcedure
	\end{algorithmic}
\end{algorithm}

\begin{algorithm}
	\caption{Series of $x$ row 1D-FFTs using FFTW interface function fftw\_plan\_many\_dft. Each row is padded to $N_{padded}$.} \label{fft_many_1d_padded}
	\begin{algorithmic}[1]
		\Procedure{1D\_ROW\_FFTS\_LOCAL\_PADDED}{\par
			\hskip\algorithmicindent\hskip\algorithmicindent $id,x,N,\mathcal{M}$}
		\INPUT
		\Statex Processor identifier, $id \in \mathbb Z_{> 0}$
		\Statex Problem size $x \in \mathbb Z_{> 0}$
		\Statex $\mathcal{M}$, Signal matrix of size $N \times N, N \in \mathbb Z_{> 0}$
		\Statex Functional performance model (speed functions) represented by,
		\Statex $\mathcal{S} = \{S_1,...,S_p\}$,
		\Statex $\mathcal{S}_i = \{(x_i[q][r],s_i[q][r])~|~i \in [1,p], q,r \in [1,m], x_i[q][r] \in \mathbb Z_{> 0}, s_i[q][r] \in \mathbb R_{> 0}\}$
		\OUTPUT
		\Statex $\mathcal{M}$, Signal matrix of size $N \times N, N \in \mathbb Z_{> 0}$
		\Statex
		\State $N_{padded} \gets \Call{Determine\_Pad\_Length}{id,x,N,\mathcal{S}}$
		\State $rank \gets 1$; $howmany \gets x$; $s \gets {N_{padded}}$;
		\State $idist \gets N_{padded}$; $odist \gets N_{padded}$; $istride \gets 1$;
		\State $ostride \gets 1$; $inembed \gets s$; $onembed \gets s$;
		\State $plan \gets \Call{fftw\_plan\_many\_dft}{rank, s, howmany,$ \par
			\hskip\algorithmicindent $\mathcal{M},inembed,istride,idist,$ \par
			\hskip\algorithmicindent $\mathcal{M},onembed,ostride,odist,$ \par
			\hskip\algorithmicindent $FFTW\_FORWARD,FFTW\_ESTIMATE$}
		\State $\Call{fftw\_execute}{plan}$
		\State $\Call{fftw\_destroy\_plan}{plan}$    
		\State \textbf{return} $\mathcal{M}$    
		\EndProcedure
	\end{algorithmic}
\end{algorithm}

Then the routine \emph{PFFT\_LIMB} is invoked to execute the basic steps 1-4 of \emph{PFFT-LB} (Line 3). These are series of row 1D-FFTs (Algorithm \ref{pfft-limb}, Lines 2-4), parallel transpose (Line 5), series of row 1D-FFTs (Lines 6-8), and parallel transpose (Line 9). 

Each processor performs the series of row 1D-FFTs locally using the routine \emph{1D\_ROW\_FFTS\_LOCAL}. The number of row 1D-FFTs performed by processor $P_i$ is given by first argument, $d_i$. The implementation of this routine using FFTW interface is shown in Algorithm \ref{fft_many_1d}.

The implementations of \emph{PFFT-FPM-PAD} are similar to those for \emph{PFFT-FPM} except that the routine \emph{1D\_ROW\_FFTS\_LOCAL\_PADDED} determines the length of the padding from the FPMs using the function \emph{Determine\_Pad\_Length} before executing the series of row 1D-FFTs.

\subsection{Shared Memory Implementations of \emph{PFFT-FPM}}

We now describe the shared-memory implementations of the routine \emph{PFFT\_LIMB} for Intel MKL FFT and FFTW-3.3.7 respectively on a Intel Haswell server containing 36 physical cores (specification is shown in Table \ref{table-Haswell-server}).

The input parameters ($p,t$) to be used during the execution of \emph{PFFT-FPM} and \emph{PFFT-FPM-PAD} are obtained from the best load-balanced configuration observed experimentally.

\subsubsection{Intel MKL FFT}

For the implementation using \emph{Intel MKL FFT}, we use two groups of 18 threads each, ($p=2,t=18$). We experimentally found this pair to be the best among the following combinations: $\{(4,9),(6,6),(9,4),(12,3)\}$, experimentally.

The routine \emph{PFFT\_LIMB\_INTEL\_MKL} shows the implementation of \emph{PFFT\_LIMB} using the FFTW interface. Lines 2-3 sets the number of threads to use during the execution of a 1D-FFT. Lines 4-8 show the execution of row 1D-FFTs by the two abstract processors (groups of 18 threads each) in parallel. Line 9 contains the fast transpose of the signal matrix. Lines 10-14 show the execution of row 1D-FFTs by the two abstract processors (groups of 18 threads each) in parallel. This is followed by fast transpose on Line 15.

The transpose routine using blocking is presented in the Appendix \ref{transpose}.

\subsubsection{FFTW}

For the implementation using \emph{FFTW-3.3.7}, we use four groups of 9 threads each, ($p=4,t=9$). We experimentally found this pair to be the best among the following combinations: $\{(2,18),(6,6),(9,4),(12,3)\}$, experimentally.

The routine \emph{PFFT\_LIMB\_FFTW} shows the implementation of \emph{PFFT\_LIMB}. Lines 2-3 sets the number of threads to use during the execution of a 1D-FFT. Lines 4-12 show the execution of row 1D-FFTs by the four abstract processors (groups of 9 threads each) in parallel. It should be noted that only thread-safe routine in FFTW is \emph{fftw\_execute}. All the other routines such an plan creation (\emph{fftw\_plan\_many\_dft}) and plan destruction (\emph{fftw\_destroy\_plan}) must be called from one thread at a time. Line 13 contains the fast transpose of the signal matrix. Lines 14-22 show the execution of row 1D-FFTs by the four abstract processors (groups of 9 threads each) in parallel. This is followed by fast transpose on Line 23.

The transpose routine using blocking is presented in the Appendix \ref{transpose}.

\section{Experimental Results and Discussion}

In this section, we present our experimental results where we present the performance improvements provided by our two model-based optimization methods, \emph{PFFT-FPM} and \emph{PFFT-FPM-PAD}, respectively. Our experimental platform is a Intel Haswell server containing 36 physical cores. Its specification is shown in Table \ref{table-Haswell-server}. 

We use two packages, FFTW-3.3.7 and Intel MKL FFT, for implementation of the algorithms. We could not optimize FFTW-2.1.5 since the implementation of series of row 1D-FFTs is quite poor using \emph{fftw\_threads} compared to the implementation of \emph{fftw\_plan\_many\_dft} in FFTW-3.3.7 and Intel MKL FFT. However, we will compare the speedups of optimized FFTW-3.3.7 and Intel MKL FFT with the unoptimized FFTW-2.1.5.

The input parameters ($p,t$) to be used during the execution of \emph{PFFT-FPM} and \emph{PFFT-FPM-PAD} are obtained from the best load-balanced configuration observed experimentally. For the implementations using \emph{FFTW-3.3.7}, we use four groups of 9 threads each, ($p=4,t=9$) since this pair performs the best among the following combinations: $\{(2,18),(6,6),(9,4),(12,3)\}$. For the implementations using \emph{Intel MKL FFT}, we use two groups of 18 threads each, ($p=2,t=18$), which was found to be the best experimentally among the following combinations: $\{(4,9),(6,6),(9,4),(12,3)\}$.

\subsection{Experimental Methodology to Build the Speed Functions} \label{experimental-methodology}

We followed the methodology described below to make sure the experimental results are reliable:
\begin{itemize}
\item The server is fully reserved and dedicated to these experiments during their execution. We also made certain that there are no drastic fluctuations in the load due to abnormal events in the server by monitoring its load continuously for a week using the tool \textit{sar}. Insignificant variation in the load was observed during this monitoring period suggesting normal and clean behavior of the server.
\item When an application is executed, it is bound to the physical cores using the \textit{numactl} tool.
\item To obtain a data point in the speed function, the application is repeatedly executed until the sample mean lies in the 95\% confidence interval and a precision of 0.025 (2.5\%) has been achieved. For this purpose, Student's t-test is used assuming that the individual observations are independent and their population follows the normal distribution. We verify the validity of these assumptions by plotting the distributions of observations.

The function $MeanUsingTtest$, shown in Algorithm \ref{mean-t-test}, describes this step. For each data point, the function is invoked, which repeatedly executes the application $app$ until one of the following three conditions is satisfied:
\begin{enumerate}
\item The maximum number of repetitions ($maxReps$) have been exceeded (Line 3).
\item The sample mean falls in the confidence interval (or the precision of measurement $eps$ has been achieved) (Lines 15-17).
\item The elapsed time of the repetitions of application execution has exceeded the maximum time allowed ($maxT$ in seconds) (Lines 18-20).
\end{enumerate}
So, for each data point, the function $MeanUsingTtest$ is invoked and the sample mean $mean$ is returned at the end of invocation. The function $Measure$ measures the execution time using the HCL's WattsUp library \cite{HCLWattsUp}. The input minimum and maximum number of repetitions, $minReps$ and $maxReps$, differ based on the problem size solved. For small problem sizes ($32 \leq n \leq 1024$), these values are set to 10000 and 100000 respectively. For medium problem sizes ($1024 < n \leq 5120$), these values are set to 100 and 1000. For large problem sizes ($n > 5120$), these values are set to 5 and 50. The values of $maxT$, $cl$, and $eps$ are respectively set to 3600, 0.95, and 0.025. If the precision of measurement is not achieved before the maximum number of repeats have been completed, we increase the number of repetitions and also the maximum elapsed time allowed. However, we observed that condition (2) is always satisfied before the other two in our experiments.
\end{itemize}

\begin{algorithm}
\caption{Function determining the mean of an experimental run using Student's t-test.}\label{mean-t-test}
\begin{algorithmic}[1]
\Procedure{MeanUsingTtest}{$app,minReps,maxReps,$ \par
$maxT,cl,accuracy,$ \par
$repsOut,clOut,etimeOut,epsOut,mean$}
\INPUT
\Statex The application to execute, $app$
\Statex The minimum number of repetitions, $minReps \in \mathbb Z_{> 0}$
\Statex The maximum number of repetitions, $maxReps \in \mathbb Z_{> 0}$
\Statex The maximum time allowed for the application to run, $maxT \in \mathbb R_{> 0}$
\Statex The required confidence level, $cl \in \mathbb R_{> 0}$
\Statex The required accuracy, $eps \in \mathbb R_{> 0}$
\OUTPUT
\Statex The number of experimental runs actually made, $repsOut \in \mathbb Z_{> 0}$
\Statex The confidence level achieved, $clOut \in \mathbb R_{> 0}$
\Statex The accuracy achieved, $epsOut \in \mathbb R_{> 0}$
\Statex The elapsed time, $etimeOut \in \mathbb R_{> 0}$
\Statex The mean, $mean \in \mathbb R_{> 0}$
\Statex
     \State $reps \gets 0$; $stop \gets 0$; $sum \gets 0$; $etime \gets 0$
     \While{($reps < maxReps$) and ($!stop$)}
        \State $st \gets \Call{measure}{TIME}$
        \State $\Call{Execute}{app}$
        \State $et \gets \Call{measure}{TIME}$
        \State $reps \gets reps + 1$
        \State $etime \gets etime + et-st$
        \State $ObjArray[reps] \gets et-st$
        \State $sum \gets sum + ObjArray[reps]$
        \If{$reps > minReps$}
           \State $clOut$ $\gets$ fabs(gsl\_cdf\_tdist\_Pinv($cl$, $reps-1$)) \par
           \hskip\algorithmicindent\hskip\algorithmicindent\hskip\algorithmicindent
           $\times$ gsl\_stats\_sd($ObjArray$, 1, $reps$) \par
           \hskip\algorithmicindent\hskip\algorithmicindent\hskip\algorithmicindent
           / sqrt($reps$)
           \If{$clOut \times \frac{reps}{sum} < eps$}
              \State $stop \gets 1$
           \EndIf
           \If{$etime > maxT$}
              \State $stop \gets 1$
           \EndIf
        \EndIf
     \EndWhile
     \State $repsOut \gets reps$; $epsOut \gets clOut \times \frac{reps}{sum}$
     \State $etimeOut \gets etime$; $mean \gets \frac{sum}{reps}$
\EndProcedure
\end{algorithmic}
\end{algorithm}

\subsection{Full Speed Functions} \label{full-speed-functions}

\begin{figure*}
	\centering
	\includegraphics[width=1.0\textwidth]{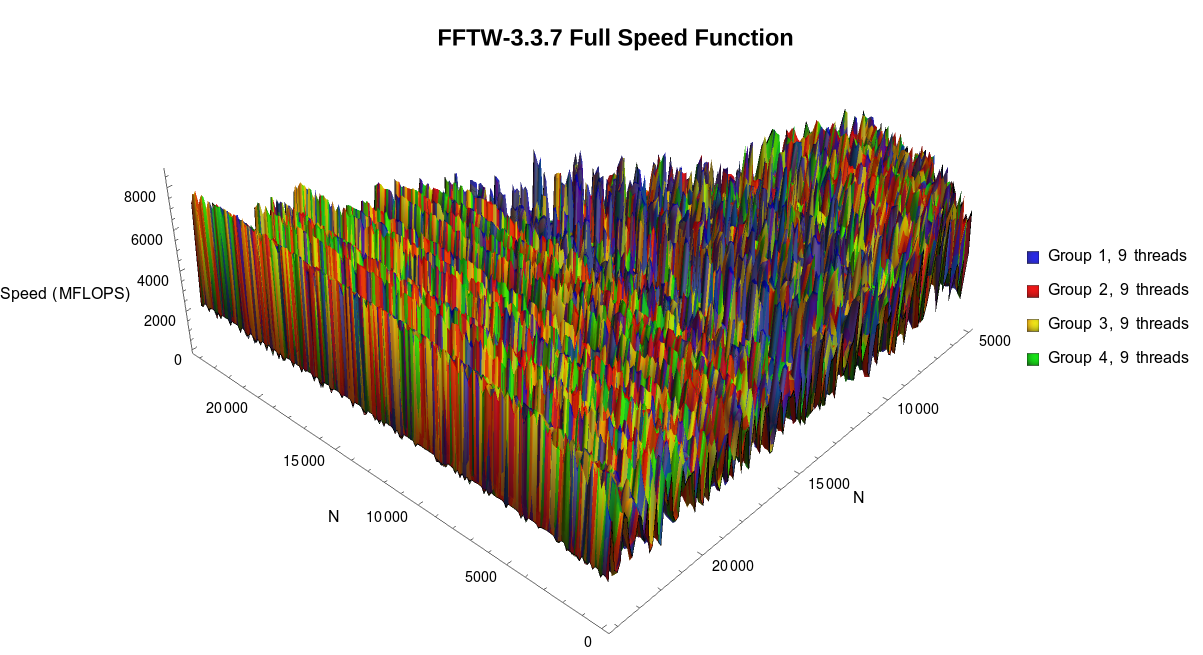}
	\caption{Full speed function of FFTW-3.3.7.}
	\label{fig:fftw337_full_speed}
\end{figure*}

\begin{figure*}
	\centering
	\includegraphics[width=1.0\textwidth]{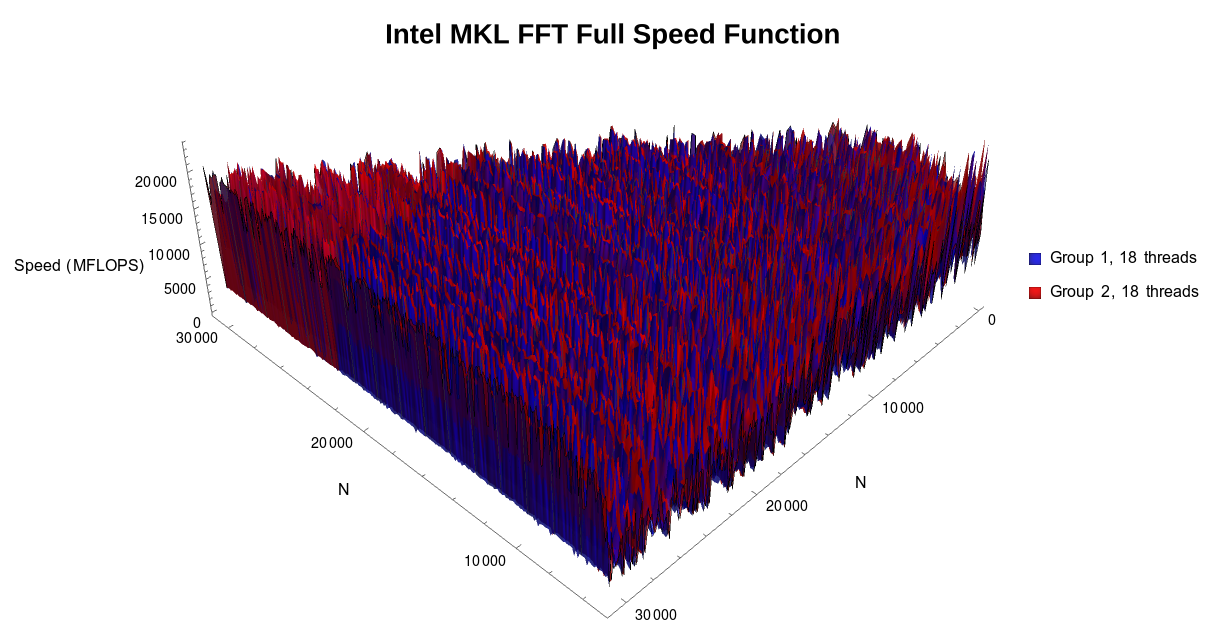}
	\caption{Full speed function of Intel MKL FFT.}
	\label{fig:imkl_full_speed}
\end{figure*}

The full speed functions constructed for Intel MKL FFT and FFTW-3.3.7 are shown in the Figures \ref{fig:fftw337_full_speed} and \ref{fig:imkl_full_speed} respectively. The inputs to the experimental methodology are the FFT application and the application parameters ($p$, $t$, $\mathcal{M}$), and the set of problem sizes. The output is the set of discrete speed functions, $\mathcal{S}=\{S_1,...S_p\}$, one for each abstract processor. The set of problem sizes $(x,y)$ used for the construction of speed functions are $\{(x,y) ~|~ 128 \le x \le y, 128 \le y \le 64000, x \bmod 128, y \bmod 128\} = \{128 \times 128, 128 \times 256, 256 \times 256,\cdots,64000 \times 64000\}$. All the abstract processors build a data point ($(x,y),s_i(x,y)$) in their speed functions simultaneously. That is, all of them execute the same problem size $x \times y$ in parallel to determine the speed $s_i(x,y)$ in their speed functions. It should be noted that for large problem sizes (for example: $\{(x,y) ~|~ 128 \le x \le 64000, y=64000$), all the data points $(x,y)$ can not be built due to main memory constraint. Therefore, the speed functions are built until permissible problem size.

The time to build the full speed functions can be quite expensive. This takes into account the fact that for each data point, statistical averaging is performed to determine its sample mean. It took around 96 hours each to build the speed functions for Intel MKL FFT and FFTW-3.3.7. However, partial speed functions \cite{Lastovetsky2011a},\cite{Lastovetsky2011b} can be built and input to the data partitioning algorithm \cite{ravilastov2017}, which would return sub-optimal data distributions (but better than load balanced solution) to be used in \emph{PFFT-FPM} and \emph{PFFT-FPM-PAD}. To build a partial speed function, data points in the neighbourhood of homogeneous distribution, $d_i=\frac{n}{p},\forall i \in [1,p]$, are constructed until the allowed user-input execution time is exceeded. We aim to research further into methods to reduce the construction times of speed functions in our future work.

\begin{figure}
	\centering
	\includegraphics[width=3.5in]{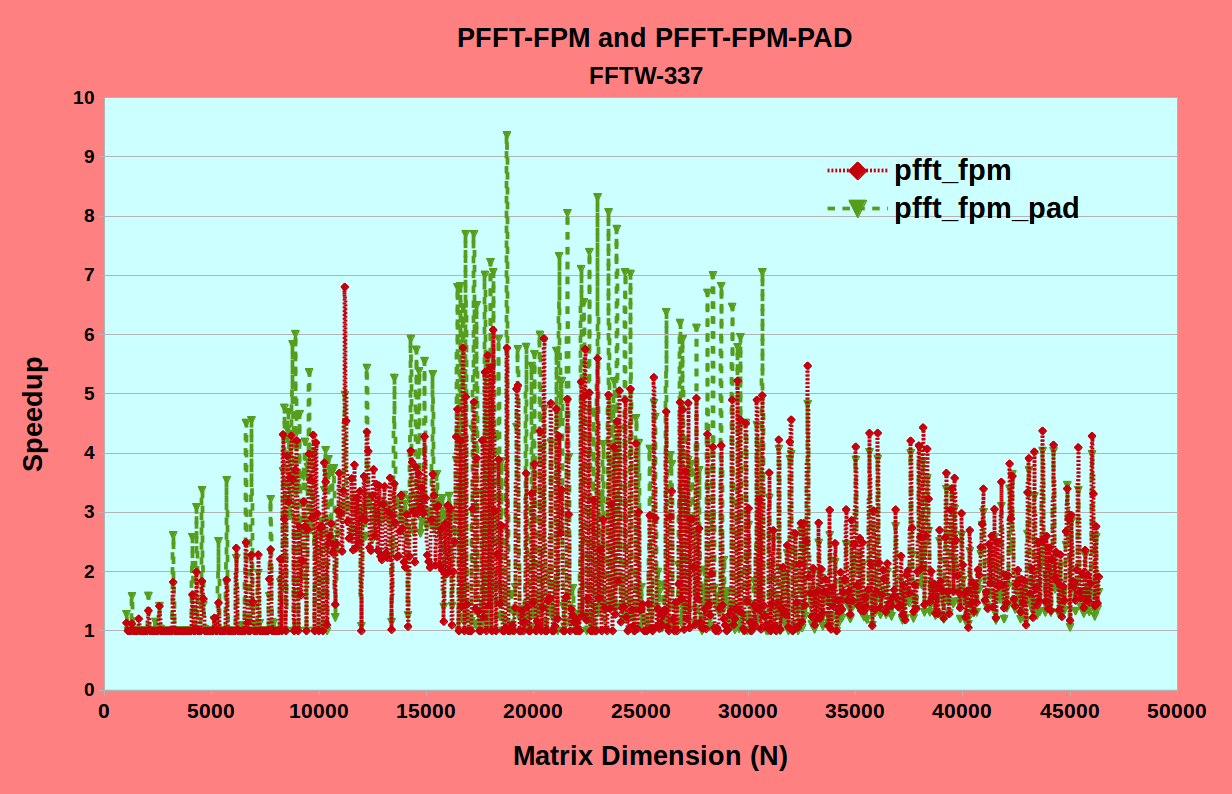}
	\caption{Speedup of \emph{PFFT-FPM} and \emph{PFFT-FPM-PAD} against the basic FFTW-3.3.7 executed using 36 threads.}
	\label{fig:fftw337_speedup}
\end{figure}

\begin{figure}
	\centering
	\includegraphics[width=3.5in]{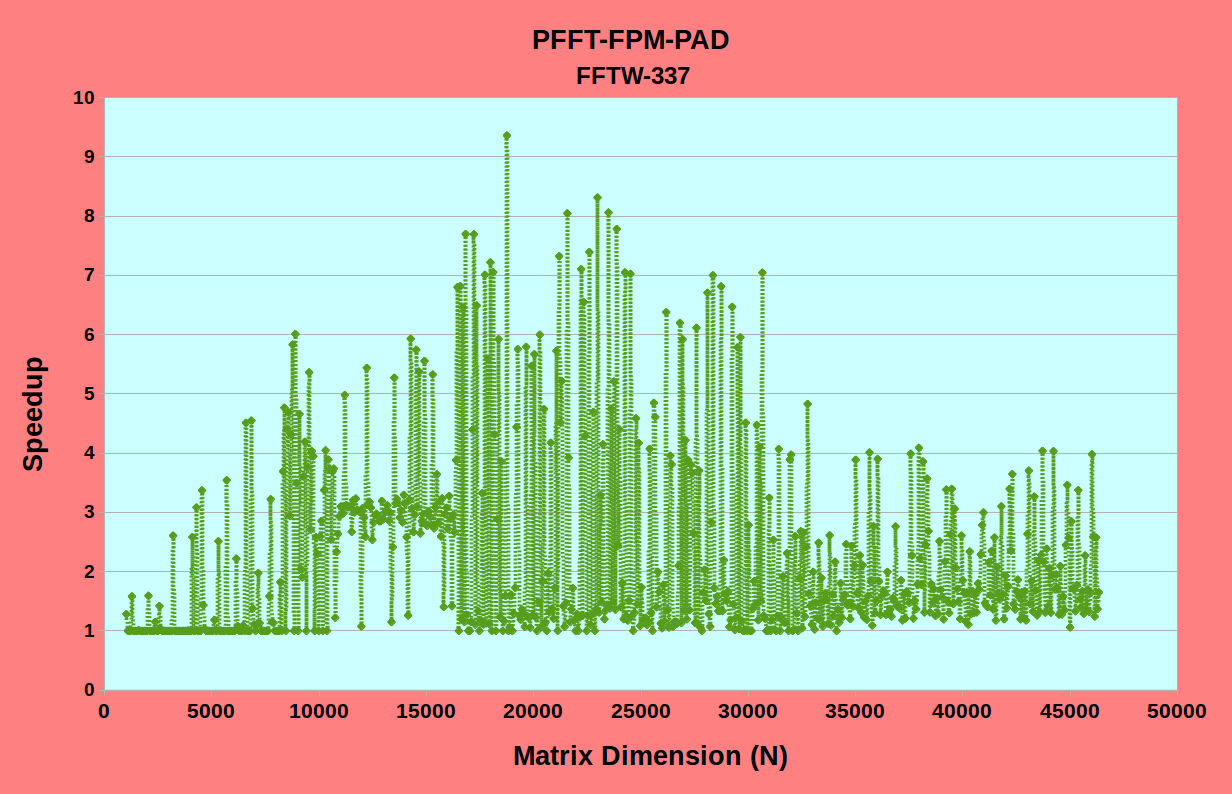}
	\caption{Speedup of \emph{PFFT-FPM-PAD} against the basic FFTW-3.3.7 executed using 36 threads.}
	\label{fig:fftw337_fpm_pad_speedup}
\end{figure}

To demonstrate the performance improvements of the solutions determined by \emph{PFFT-FPM} and \emph{PFFT-FPM-PAD}, we report the average and maximum speedups over to the basic FFT versions (that employ one groups of 36 threads in their execution). For \emph{PFFT-FPM}, the speedup is calculated as follows: $\text{Speedup} = \frac{t_{basic}}{t_{pfft-fpm}}$, where $t_{basic}$ is the execution time obtained using the basic FFT version (Intel MKL FFT or FFTW-3.3.7) and $t_{pfft-fpm}$ is the execution time obtained using \emph{PFFT-FPM}. For \emph{PFFT-FPM-PAD}, the speedup is calculated as follows: $\text{Speedup} = \frac{t_{basic}}{t_{pfft-fpm-pad}}$, where $t_{pfft-fpm-pad}$ is the execution time obtained using \emph{PFFT-FPM-PAD}.

\subsection{\emph{PFFT-FPM} and \emph{PFFT-FPM-PAD} using FFTW-3.3.7}

Figure \ref{fig:fftw337_speedup} shows the speedups of \emph{PFFT-FPM} and \emph{PFFT-FPM-PAD} over basic FFTW-3.3.7 where the 2D-DFT is computed using one group consisting of 36 threads. Each data point in the speed functions involves a complex 2D-DFT of size $N \times N$. Figure \ref{fig:fftw337_fpm_pad_speedup} shows the speedup of \emph{PFFT-FPM-PAD} for problem sizes where performance has been improved. The average and maximum performance improvements are 2x and 9.4x respectively.

\begin{figure}
	\centering
	\includegraphics[width=3.5in]{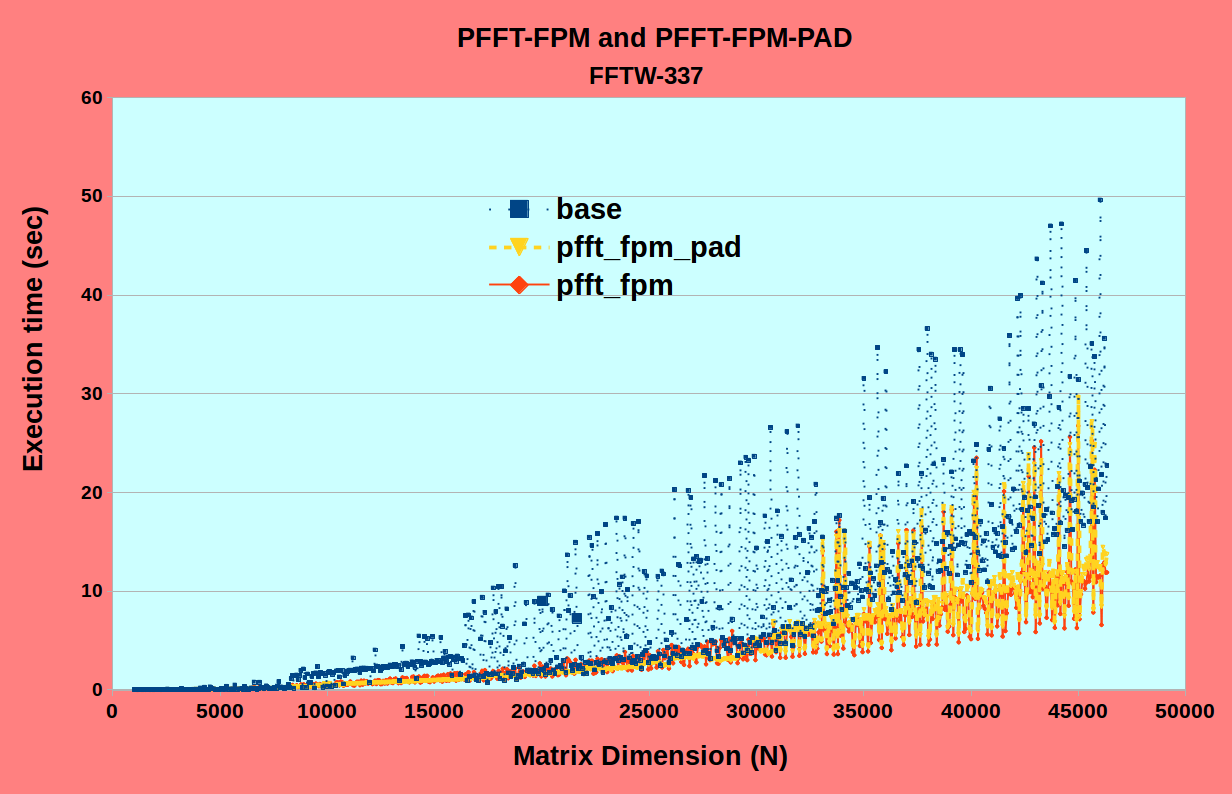}
	\caption{Execution times of \emph{PFFT-FPM} and \emph{PFFT-FPM-PAD} against the basic FFTW-3.3.7 executed using 36 threads.}
	\label{fig:fftw337_fpm_etimes}
\end{figure}

\begin{figure}
	\centering
	\includegraphics[width=3.5in]{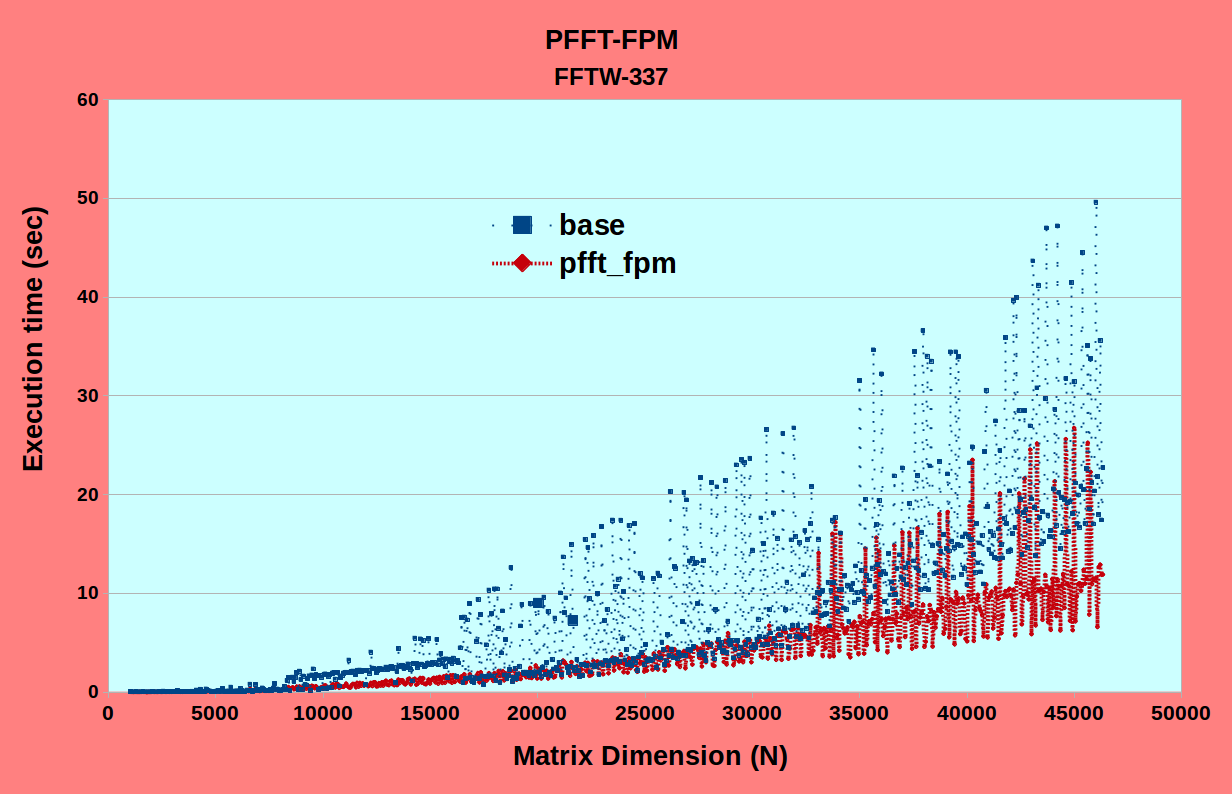}
	\caption{Execution times of \emph{PFFT-FPM} against the basic FFTW-3.3.7.}
	\label{fig:fftw337_fpm_etimes3}
\end{figure}

\begin{figure}
	\centering
	\includegraphics[width=3.5in]{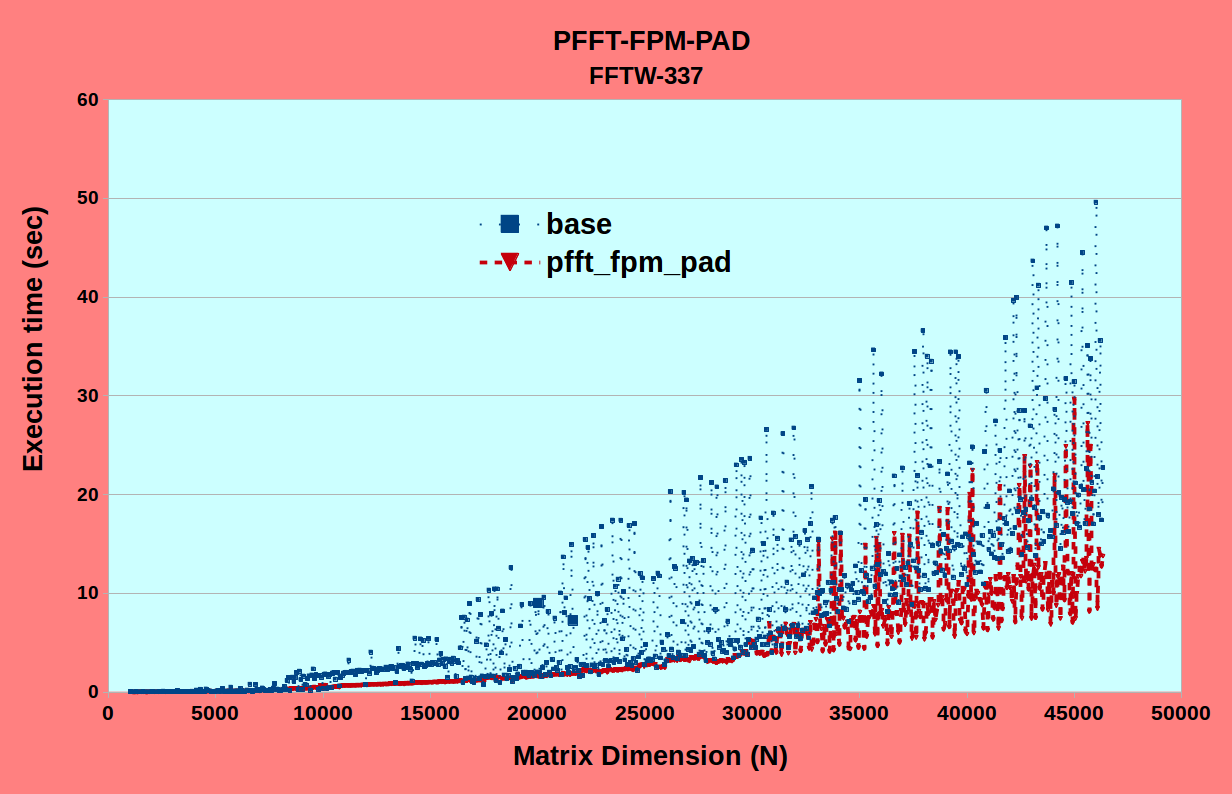}
	\caption{Execution times of \emph{PFFT-FPM-PAD} against the basic FFTW-3.3.7.}
	\label{fig:fftw337_fpm_etimes2}
\end{figure}

Figure \ref{fig:fftw337_fpm_etimes} shows the execution times of \emph{PFFT-FPM} and \emph{PFFT-FPM-PAD} versus basic FFTW-3.3.7. Figure \ref{fig:fftw337_fpm_etimes3} shows the execution times of \emph{PFFT-FPM} only versus basic FFTW-3.3.7. Figure \ref{fig:fftw337_fpm_etimes2} shows the execution times of \emph{PFFT-FPM-PAD} only versus basic FFTW-3.3.7.

For problem sizes in the range ($N > 33000$), while the speedups are still quite good (6x for FFTW-3.3.7), major variations still remain.

\subsection{\emph{PFFT-FPM} and \emph{PFFT-FPM-PAD} using Intel MKL FFT}

\begin{figure}
	\centering
	\includegraphics[width=3.5in]{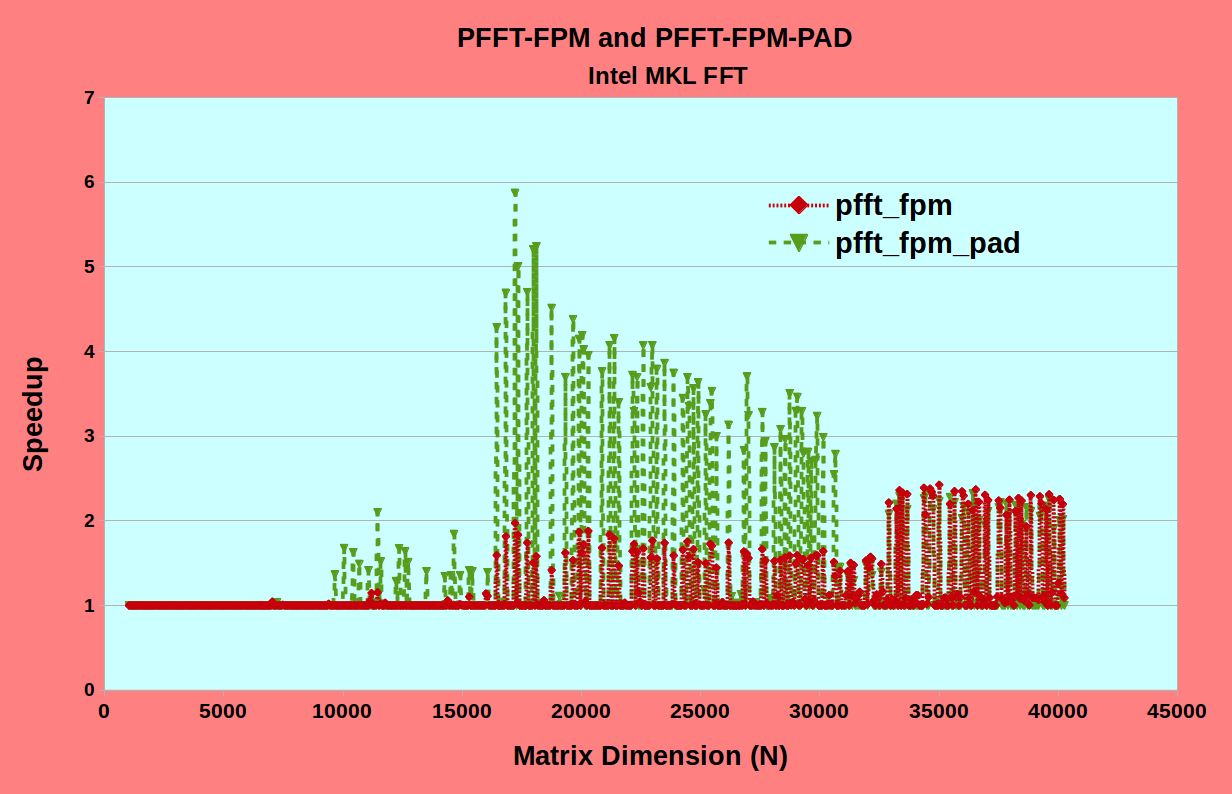}
	\caption{Speedups of \emph{PFFT-FPM} and \emph{PFFT-FPM-PAD} against the basic Intel MKL FFT executed using 36 threads.}
	\label{fig:imkl_speedup}
\end{figure}

\begin{figure}
	\centering
	\includegraphics[width=3.5in]{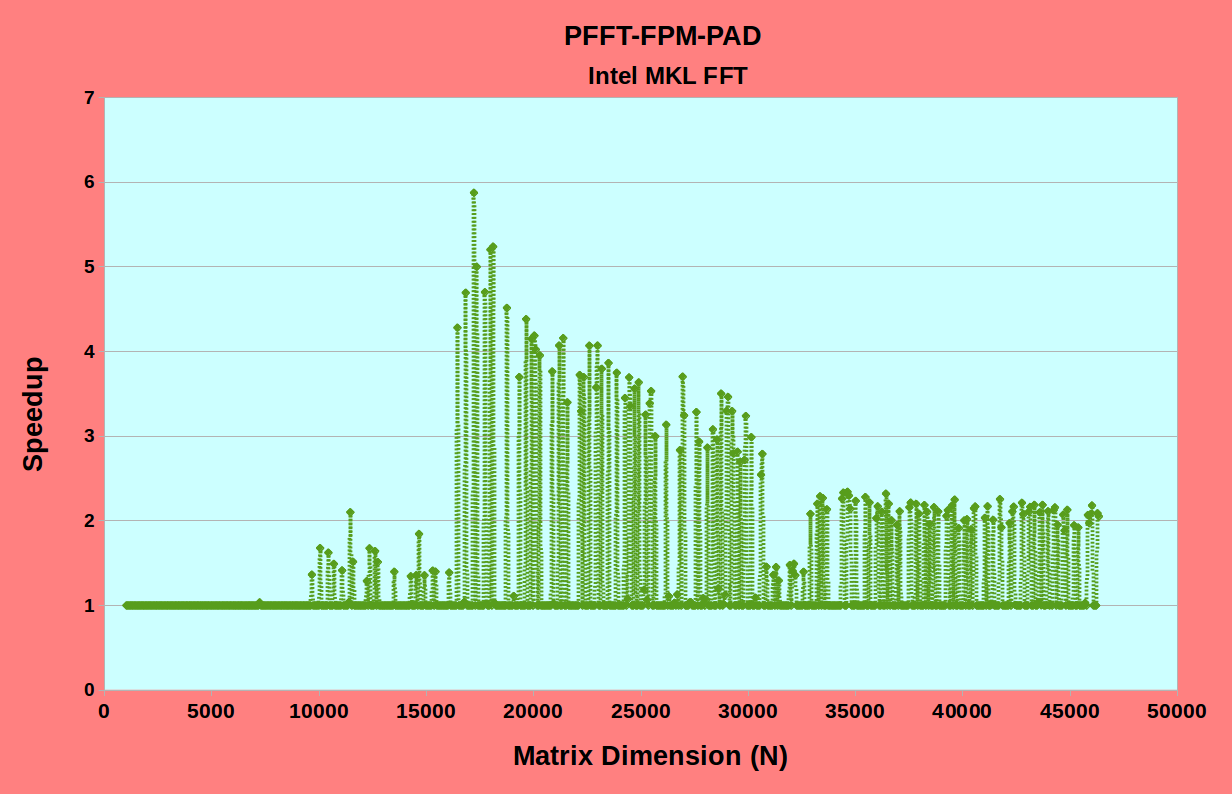}
	\caption{Speedup of \emph{PFFT-FPM-PAD} against the basic Intel MKL FFT executed using 36 threads.}
	\label{fig:imkl_fpm_pad_speedup}
\end{figure}

Figure \ref{fig:imkl_speedup} compares the speedups \emph{PFFT-FPM} and \emph{PFFT-FPM-PAD} over basic Intel MKL FFT where the 2D-DFT is computed using one group consisting of 36 threads. Figure \ref{fig:imkl_fpm_pad_speedup} shows the speedups of \emph{PFFT-FPM-PAD} for problem sizes where performance has been improved. The average and maximum speedups are 1.4x and 5.9x respectively.

\begin{figure}
	\centering
	\includegraphics[width=3.5in]{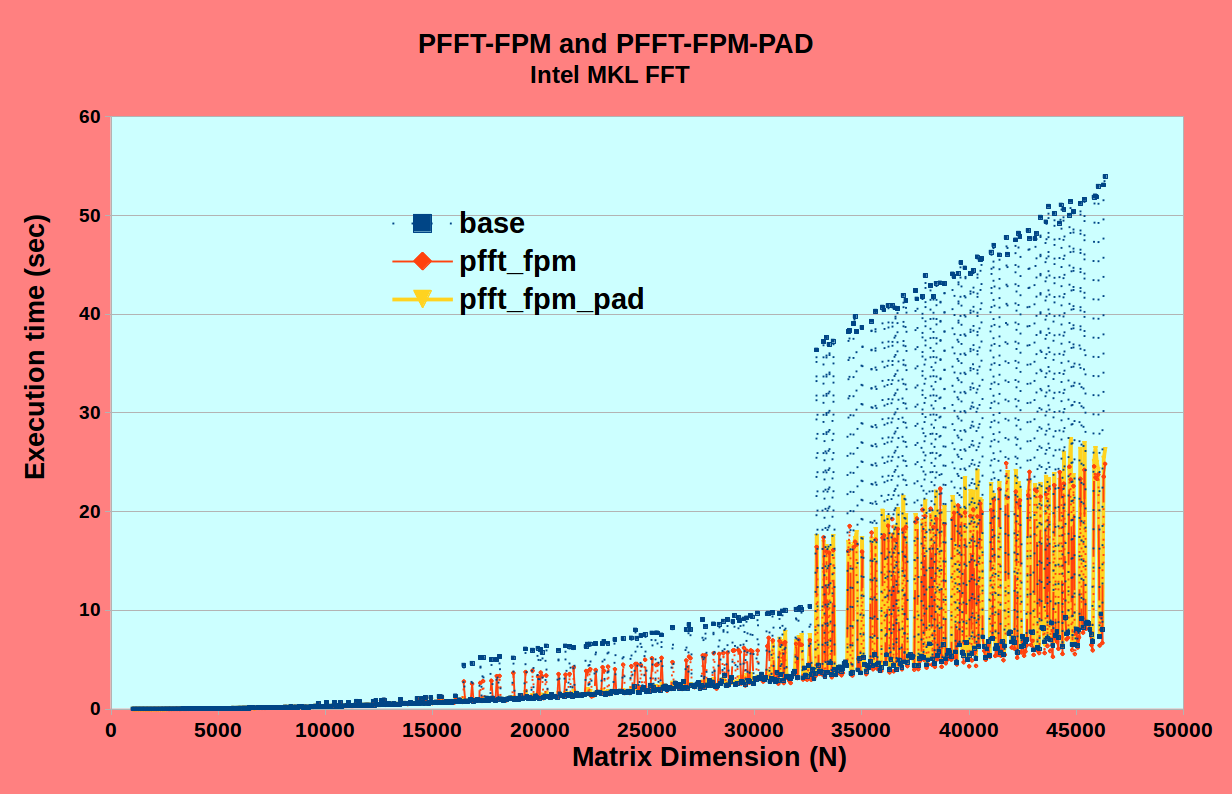}
	\caption{Execution times of \emph{PFFT-FPM} and \emph{PFFT-FPM-PAD} against the basic Intel MKL FFT executed using 36 threads.}
	\label{fig:imkl_fpm_pad_etimes}
\end{figure}

\begin{figure}
	\centering
	\includegraphics[width=3.5in]{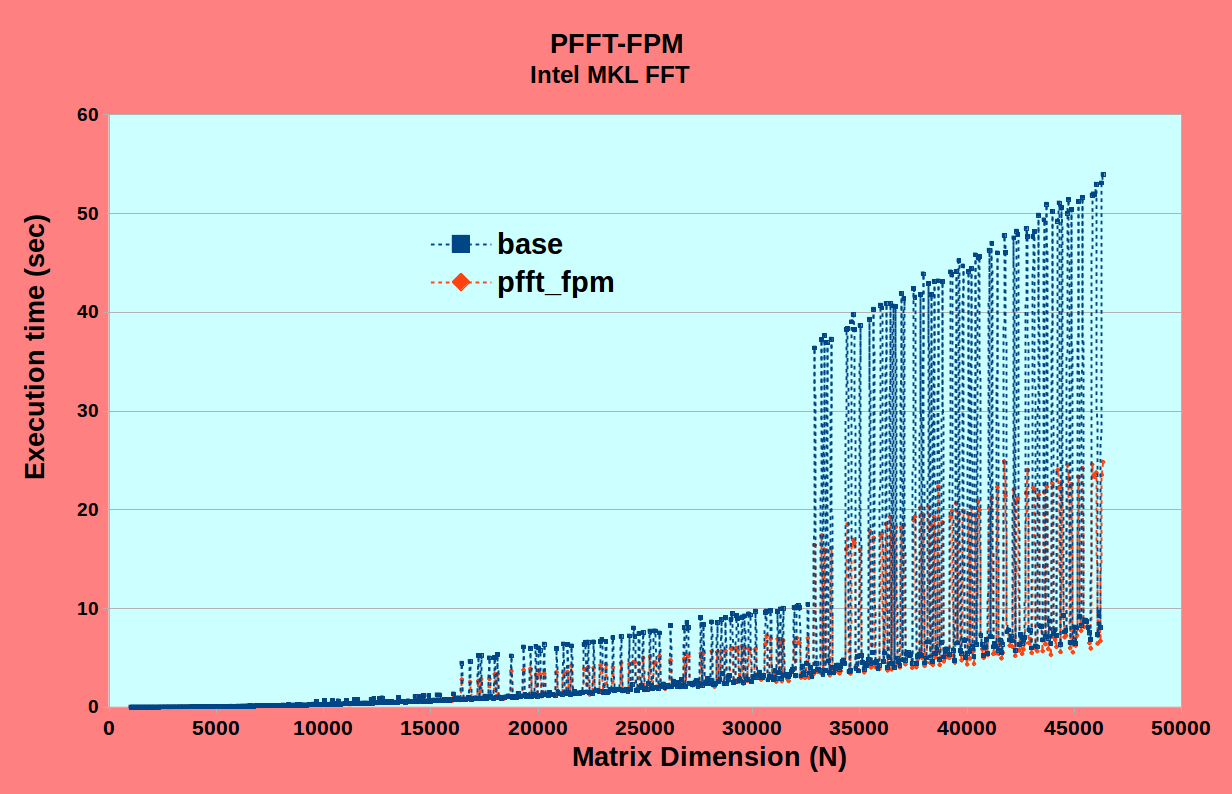}
	\caption{Execution times of \emph{PFFT-FPM} against the basic Intel MKL FFT.}
	\label{fig:imkl_fpm_pad_etimes2}
\end{figure}

\begin{figure}
	\centering
	\includegraphics[width=3.5in]{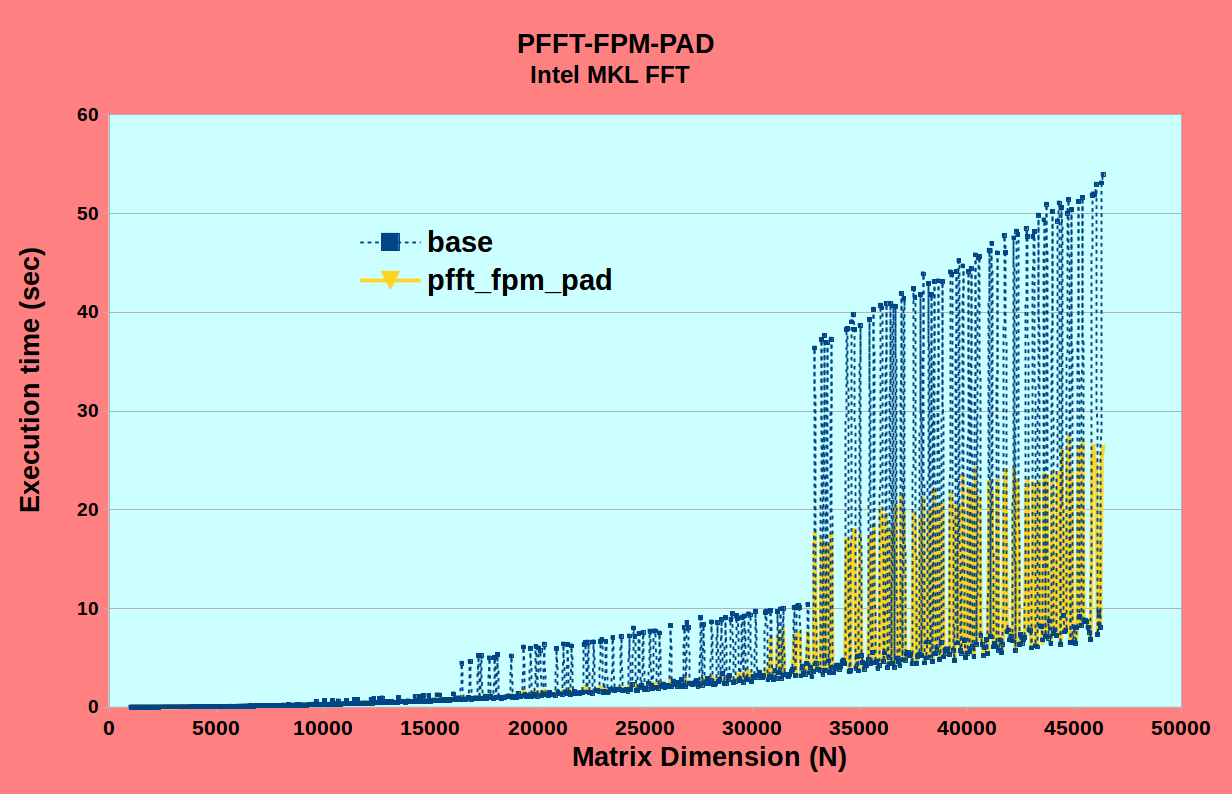}
	\caption{Execution times of \emph{PFFT-FPM-PAD} against the basic Intel MKL FFT.}
	\label{fig:imkl_fpm_pad_etimes3}
\end{figure}

Figure \ref{fig:imkl_fpm_pad_etimes} shows the execution times of \emph{PFFT-FPM} and \emph{PFFT-FPM-PAD} versus basic Intel MKL FFT.  Figure \ref{fig:imkl_fpm_pad_etimes2} shows the execution times of \emph{PFFT-FPM} only versus basic Intel MKL FFT. Figure \ref{fig:imkl_fpm_pad_etimes3} shows the execution times of \emph{PFFT-FPM-PAD} only versus basic Intel MKL FFT. 

For problem sizes in the range ($N > 33000$), while the speedups are still quite good (2x for Intel MKL FFT), the variations are still quite significant. 

\subsection{Optimized FFTW-3.3.7 and Intel MKL FFT versus Unoptimized FFTW-2.1.5}

Finally, we compare how the optimized FFTW-3.3.7 and Intel MKL FFT using \emph{PFFT-FPM-PAD} fares with respect to unoptimized FFTW-2.1.5. 

\begin{figure}
	\centering
	\includegraphics[width=3.5in]{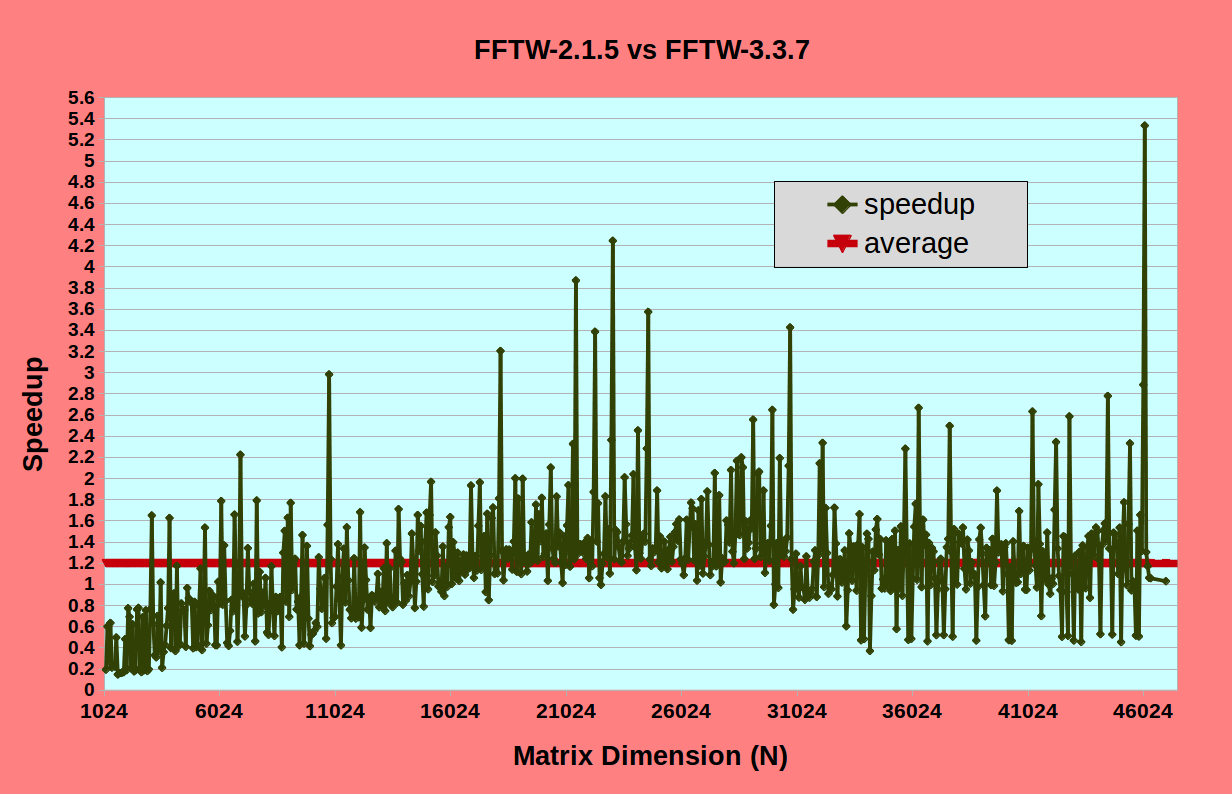}
	\caption{Speedup of optimized FFTW-3.3.7 (using \emph{PFFT-FPM-PAD}) over unoptimized FFTW-2.1.5. The average speedup is 1.2x.}
	\label{fig:fftw215_fftw337_fpmpad}
\end{figure}

Figure \ref{fig:fftw215_fftw337_fpmpad} shows the speedup of FFTW-3.3.7 using \emph{PFFT-FPM-PAD} versus unoptimized FFTW-2.1.5. One can see that in the range of problem sizes ($N < 15000$), FFTW-2.1.5 performs better than FFTW-3.3.7. There are few problem sizes in the range ($N > 30000$) again where it is better. The average performances of FFTW-3.3.7 and FFTW-2.1.5 are 7297 MFLOPs and 7033 MFLOPs respectively. The average speedup of FFTW-3.3.7 over FFTW-2.1.5 is 1.2x. Most importantly, our optimizations have improved the average performance of FFTW-3.3.7 over FFTW-2.1.5 by 42\%. 

\begin{figure}
	\centering
	\includegraphics[width=3.5in]{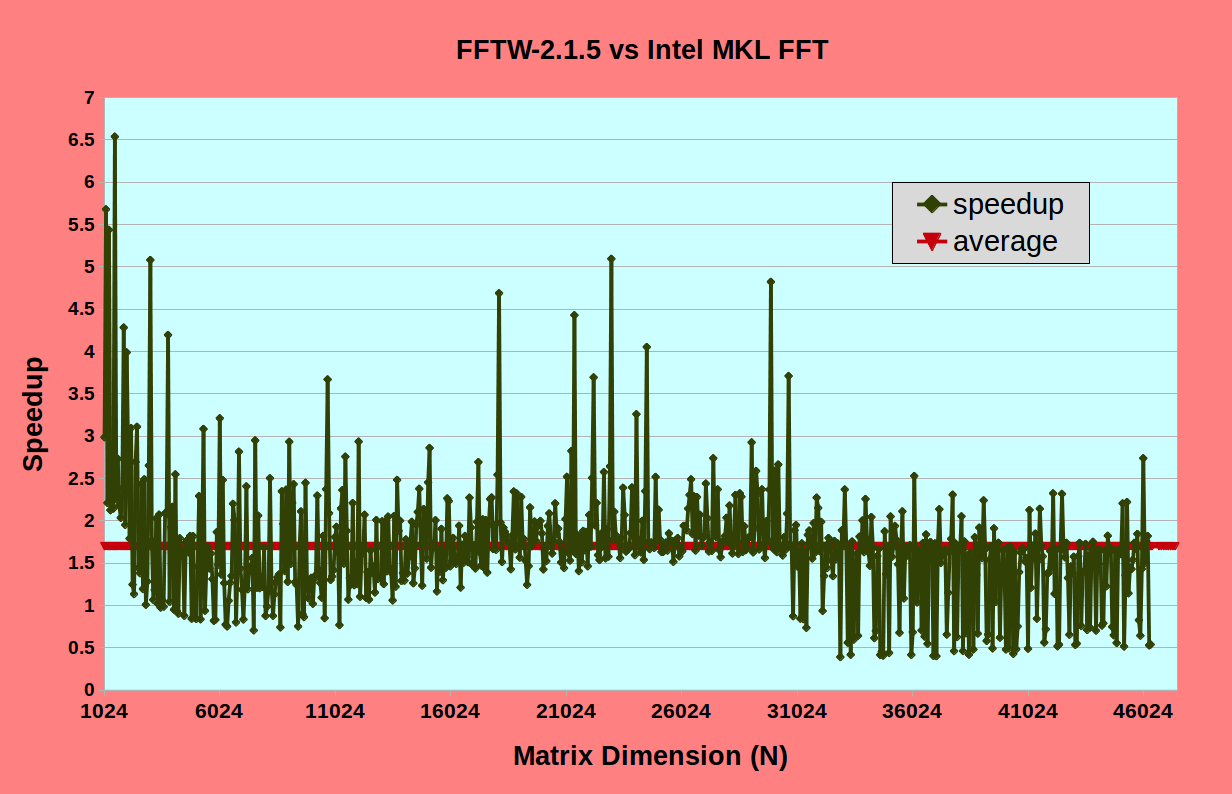}
	\caption{Speedup of optimized Intel MKL FFT (using \emph{PFFT-FPM-PAD}) over unoptimized FFTW-2.1.5. The average speedup is 1.7x.}
	\label{fig:fftw215_imkl_fpmpad}
\end{figure}

Figure \ref{fig:fftw215_imkl_fpmpad} shows the speedup of Intel MKL FFT using \emph{PFFT-FPM-PAD} versus unoptimized FFTW-2.1.5. The average performances of Intel MKL FFT and FFTW-2.1.5 are 11170 MFLOPs and 7033 MFLOPs respectively (Intel MKL FFT being 60\% better). However, there are around 91 problem sizes (majority of them closer to the end of the figure) where FFTW-2.1.5 exhibits better performance than Intel MKL FFT. Most importantly, our optimizations have improved the average performance of Intel MKL FFT over FFTW-2.1.5 by 24\% (over and above the 36\% of unoptimized Intel MKL FFT). The average speedup of FFTW-3.3.7 over FFTW-2.1.5 is 1.7x.

\subsection{Summary}

We summarize the results below:

\begin{itemize}
	\item For problem sizes in the range ($0 < N \le 10000$), the speedups provided by \emph{PFFT-FPM} and \emph{PFFT-FPM-PAD} for Intel MKL FFT are not significant. This is because the variations (performance drops) are not remarkable.
	\item For problem sizes in the range ($10000 < N \le 33000$), the speedups are tremendous. \\
	For FFTW-3.3.7, the average and maximum speedups provided by \emph{PFFT-FPM} are 2.7x and 6.8x respectively and those provided by \emph{PFFT-FPM-PAD} are 3x and 9.4x respectively. \\
	For Intel MKL FFT, the average and maximum speedups provided by \emph{PFFT-FPM} are 1.4x and 2x respectively and those provided by \emph{PFFT-FPM-PAD} are 2.7x and 5.9x respectively.
	The variations (performance drops) have been virtually completely removed.
	\item For problem sizes in the range ($N > 33000$), the speedups are good but not as excellent and major variations still remain. The variations are more severe for Intel MKL FFT. We aim to find solutions to remove them in our future work.
	\item The average speeds/performances of \emph{PFFT-FPM} using FFTW-3.3.7 and Intel MKL FFT are 7041 MFLOPs and 10818 MFLOPs respectively. So, Intel MKL FFT is on an average 54\% better than FFTW-3.3.7. However, there are 135 problem sizes (out of 700) where FFTW-3.3.7 outperforms Intel MKL FFT. The average speeds/performances of \emph{PFFT-FPM-PAD} using FFTW-3.3.7 and Intel MKL FFT are 7297 MFLOPs and 11170 MFLOPs respectively. There are 81 problem sizes (out of 700) where FFTW-3.3.7 outperforms Intel MKL FFT. So, Intel MKL FFT is on an average 53\% better than FFTW-3.3.7.
	\item The optimized FFTW-3.3.7 and Intel MKL FFT using \emph{PFFT-FPM-PAD} demonstrate average performance improvements of 42\% and 24\% respectively over FFTW-2.1.5. However, there are problem sizes where FFTW-2.1.5 still performs better than FFTW-3.3.7 and Intel MKL FFT. This will be the subject of our future research.
\end{itemize}

\section{Related Work}

In this section, we review parallel solutions proposed for performance optimization of FFT on both homogeneous and heterogeneous platforms. We survey load-balancing algorithms employed for performance optimization of scientific applications on modern multicore platforms. Finally, we present an overview of the latest efforts addressing the variations using load imbalancing algorithms on modern multicore platforms.

\subsection{Parallel FFT solutions for homogeneous and heterogeneous platforms}

There are several works that present parallel FFTs for distributed memory architectures. Averbuch et al. \cite{averbuch1998} present a parallel version of the Cooley–Tukey FFT algorithm for MIMD multiprocessors and demonstrate efficiency of 90\% on a message-passing IBM SP2 computer.

Dmitruk et al. \cite{dmitruk2001} use a 1D domain decomposition algorithm for performance improvement of 3D real FFT. They present techniques for reducing the cost of communications in the communication-intensive transpose operation of their algorithm.

We review few research works that have proposed optimized FFT implementations for GPU platforms. Chen et al. \cite{chen2010} present optimized FFT implementations for GPU clusters. Gu et al. \cite{gu2011} propose out-of-card implementations for 1D, 2D, and 3D FFTs on GPUs. Wu et al. \cite{wu2014} present optimized multi-dimensional FFT implementations on CPU–GPU heterogeneous platforms where the input signal matrix is too large to fit in the GPU global memory. Naik et al. \cite{naik2015analysis} demonstrate good performance improvement of FFT on their heterogeneous cluster compared to a homogeneous cluster.

\subsection{Parallel FFT Libraries}

The Fastest Fourier Transform in the West (FFTW) \cite{FFTW}, \cite{PFFTW} is a software library for computing discrete Fourier transforms (DFTs). It provides routines utilizing threads for parallel one- and multi-dimensional transforms of both real and complex data, and multi-dimensional transforms of real and complex data for parallel machines supporting MPI.

Pekurovsky et al. \cite{pekurovsky2012p3dfft} present a library P3DFFT, which computes fast Fourier transforms (FFTs) in three dimensions by using two-dimensional domain decomposition. Li et al. \cite{li20102decomp} provides an to perform three-dimensional distributed FFTs using MPI. OpenFFT \cite{Duy2014} is an open source parallel package for computing multi-dimensional Fast Fourier Transforms (3-D and 4-D FFTs) of both real and complex numbers of arbitrary input size.

The Intel Math Kernel library (Intel MKL) \cite{IntelMKLFFT} provides an interface for computing a discrete Fourier transform  in one, two, or three dimensions with support for mixed radices. It provides DFT routines for single-processor or shared-memory systems, and for distributed-memory architectures.

\subsection{Load balancing algorithms for performance optimization on multicore platforms}

Load balancing is a widely used method for performance optimization of scientific applications on parallel platforms. There are several classifications of it: static or dynamic, centralized or distributed, and synchronous or asynchronous.

Static algorithms use \textit{a priori} information about the parallel application and platform \cite{Lastovetsky2007}, \cite{ogata2008efficient}. They are particularly useful for applications where data locality is important because they do not require data redistribution. However, these algorithms are may be unsuitable for non-dedicated platforms, where load changes with time.. 

Dynamic algorithms balance the load by moving fine-grained tasks between processors during the execution \cite{linderman2008merge}, \cite{quintana2009solving}, \cite{augonnet2009automatic}. They often use static partitioning for their initial step due to its provably near-optimal communication cost, bounded tiny load imbalance, and lesser scheduling overhead. 

In the non-centralized load balancing algorithms, at some point of computation, each processor find neighbours that are less loaded than itself and redistributes data between them \cite{cybenko1989dynamic}, \cite{bahi2005dynamic}. In centralized algorithms, there is a centralized load balancer that decides when to distribute data based on global load information \cite{Carino2008}, \cite{Martinez2011}. 

The synchronous algorithm means that for each processor to balance its load at time $t + 1$, a processor needs to have the load of its neighbor at time $t$ \cite{bahi2005synchronous}. In other words, there is time-synchronization between all processors. In an asynchronous algorithm, the time synchronization is absent \cite{liu2016asynchronous}.

The most advanced load balancing algorithms use functional performance models (FPMs), which are application-specific and represent the speed of a processor by continuous function of problem size but satisfying some assumptions on its shape \cite{Lastovetsky2004},\cite{Lastovetsky2007}. These FPMs capture accurately the real-life behaviour of applications executing on nodes consisting of uniprocessors (single-core CPUs).

\subsection{Load imbalancing algorithms for performance optimization on multicore platforms}

Lastovetsky et al. \cite{Lastovetsky2015a}, \cite{Lastovetsky2015b} study the variations in performance profile for a real-life data-parallel scientific application, Multidimensional Positive Definite Advection Transport Algorithm (MPDATA), on a Xeon Phi co-processor. This is the first work where the load-imbalancing technique is applied to distribute the workload unevenly minimizing the computation time of its parallel execution. However, no general partitioning algorithm is proposed in this work.

Lastovetsky et al.  \cite{ravilastov2017}, Reddy et al. \cite{Reddy2016}, and Khaleghzadeh et al. \cite{hamid2018} are theoretical works that present novel data partitioning algorithms for minimization of time and energy of computations for the most general performance and energy profiles of data-parallel applications executing on homogeneous and heterogeneous multicore clusters.

In this paper, we present novel model-based methods for performance optimization of a real-life multithreaded application (2D-DFT) on multicore processors.
   
\section{Conclusion}

Code modernization experts are engaged in a perpetual battle to keep their codes up-to-date with the ever-changing hardware landscape by porting and tuning them to extract the utmost performance from the current hardware platforms. They commonly use roofline model to gauge the performance gains accrued from incremental optimizations towards achieving the theoretical peak performance of a processor. However, since hardware platforms are changing at a rapid pace, this practice of incremental nodal optimization using architecture-specific techniques can be retrogressive with two typical symptoms. First, it can fall prey to Red Queen Principle where one spends several man-years putting extensive optimizations only in the long run to stay in the same place where one started. Second, it is very likely that an open source package with portable optimizations may exhibit better performance for some problem sizes and better average performance overall than a heavily optimized vendor package. 

In this paper, we expounded this insight using multithreaded Fast Fourier transforms provided in three highly optimized packages, FFTW-2.1.5, FFTW-3.3.7, and Intel MKL FFT. Then, we proposed two novel model-based optimization methods, \emph{PFFT-FPM} and \emph{PFFT-FPM-PAD}, that employ parallel computing based on advanced functional performance models and are therefore highly portable. They compute 2D-DFT of a complex signal matrix of size $N \times N$ using $p$ abstract processors. Both the algorithms take as inputs, discrete 3D functions of performance against problem size of the processors and output the transformed signal matrix.

We performed our experiments on a modern Intel Haswell multicore server consisting of two processors of 18 physical cores each. The average and maximum speedups observed for \emph{PFFT-FPM} using \emph{FFTW-3.3.7} are 1.9x and 6.8x respectively and the average and maximum speedups observed using \emph{Intel MKL FFT} are 1.3x and 2x respectively. The average and maximum speedups observed for \emph{PFFT-FPM-PAD} using \emph{FFTW-3.3.7} are 2x and 9.4x respectively and the average and maximum speedups observed using \emph{Intel MKL FFT} are 1.4x and 5.9x respectively. We showed that using our optimization methods improves the average performance of FFTW-3.3.7 over the unoptimized FFTW-2.1.5 by 42\% and the average performance of Intel MKL FFT over the unoptimized FFTW-2.1.5 by 24\% (over and above the 36\% of unoptimized Intel MKL FFT).

The software implementations of the algorithms presented in this paper can be found at \cite{hclfft}.

In our future work, we plan to extend our algorithms for fast computation of 3D-DFT. We would also develop extensions of them for homogeneous and heterogeneous clusters of multicore nodes.

\ifCLASSOPTIONcompsoc
  \section*{Acknowledgments}
\else
  \section*{Acknowledgment}
\fi

This publication has emanated from research conducted with the financial support of Science Foundation Ireland (SFI) under Grant Number 14/IA/2474.

\bibliographystyle{IEEEtran}
\bibliography{IEEEabrv,paper}

\begin{thebibliography}{10}
\providecommand{\url}[1]{#1}
\csname url@samestyle\endcsname
\providecommand{\newblock}{\relax}
\providecommand{\bibinfo}[2]{#2}
\providecommand{\BIBentrySTDinterwordspacing}{\spaceskip=0pt\relax}
\providecommand{\BIBentryALTinterwordstretchfactor}{4}
\providecommand{\BIBentryALTinterwordspacing}{\spaceskip=\fontdimen2\font plus
\BIBentryALTinterwordstretchfactor\fontdimen3\font minus
  \fontdimen4\font\relax}
\providecommand{\BIBforeignlanguage}[2]{{%
\expandafter\ifx\csname l@#1\endcsname\relax
\typeout{** WARNING: IEEEtran.bst: No hyphenation pattern has been}%
\typeout{** loaded for the language `#1'. Using the pattern for}%
\typeout{** the default language instead.}%
\else
\language=\csname l@#1\endcsname
\fi
#2}}
\providecommand{\BIBdecl}{\relax}
\BIBdecl

\bibitem{ishizaka2003cache}
K.~Ishizaka, M.~Obata, and H.~Kasahara, ``Cache optimization for coarse grain
  task parallel processing using inter-array padding,'' in \emph{International
  Workshop on Languages and Compilers for Parallel Computing}.\hskip 1em plus
  0.5em minus 0.4em\relax Springer, 2003, pp. 64--76.

\bibitem{Zhao2007}
P.~Zhao, S.~Cui, Y.~Gao, R.~Silvera, and J.~N. Amaral, ``Forma: A framework for
  safe automatic array reshaping,'' \emph{ACM Trans. Program. Lang. Syst.},
  vol.~30, no.~1, Nov. 2007.

\bibitem{Hong2016}
C.~Hong, W.~Bao, A.~Cohen, S.~Krishnamoorthy, L.-N. Pouchet, F.~Rastello,
  J.~Ramanujam, and P.~Sadayappan, ``Effective padding of multidimensional
  arrays to avoid cache conflict misses,'' in \emph{Proceedings of the 37th ACM
  SIGPLAN Conference on Programming Language Design and Implementation}, ser.
  PLDI '16.\hskip 1em plus 0.5em minus 0.4em\relax ACM, 2016, pp. 129--144.

\bibitem{Jiang2017}
P.~Jiang and G.~Agrawal, ``Efficient {SIMD} and {MIMD} parallelization of
  hash-based aggregation by conflict mitigation,'' in \emph{Proceedings of the
  International Conference on Supercomputing}, ser. ICS '17.\hskip 1em plus
  0.5em minus 0.4em\relax ACM, 2017, pp. 24:1--24:11.

\bibitem{ravilastov2017}
A.~Lastovetsky and R.~Reddy, ``New model-based methods and algorithms for
  performance and energy optimization of data parallel applications on
  homogeneous multicore clusters,'' \emph{IEEE Transactions on Parallel and
  Distributed Systems}, vol.~28, no.~4, pp. 1119--1133, 2017.

\bibitem{hamid2018}
H.~Khaleghzadeh, R.~Reddy, and A.~Lastovetsky, ``A novel data-partitioning
  algorithm for performance optimization of {Data-Parallel} applications on
  heterogeneous {HPC} platforms,'' \emph{IEEE Transactions on Parallel and
  Distributed Systems}, 2018.

\bibitem{HCLWattsUp}
\BIBentryALTinterwordspacing
R.~Reddy and A.~L. Lastovetsky, ``{HCLWattsUp}: {API} for power and energy
  measurements using {WattsUp Pro Meter},'' 2016. [Online]. Available:
  \url{http://git.ucd.ie/hcl/hclwattsup}
\BIBentrySTDinterwordspacing

\bibitem{Lastovetsky2011a}
D.~Clarke, A.~Lastovetsky, and V.~Rychkov, ``Dynamic load balancing of parallel
  computational iterative routines on highly heterogeneous {HPC} platforms,''
  \emph{Parallel Processing Letters}, vol.~21, pp. 195--217, 06/2011 2011.

\bibitem{Lastovetsky2011b}
A.~Lastovetsky, R.~Reddy, V.~Rychkov, and D.~Clarke, ``Design and
  implementation of self-adaptable parallel algorithms for scientific computing
  on highly heterogeneous {HPC} platforms,'' \emph{arXiv preprint
  arXiv:1109.3074}, 2011.

\bibitem{averbuch1998}
A.~Averbuch and E.~Gabber, ``Portable parallel {FFT} for {MIMD}
  multiprocessors,'' \emph{Concurrency: Practice and Experience}, vol.~10,
  no.~8, 1998.

\bibitem{dmitruk2001}
P.~Dmitruk, L.-P. Wang, W.~Matthaeus, R.~Zhang, and D.~Seckel, ``Scalable
  parallel {FFT} for spectral simulations on a {Beowulf} cluster,''
  \emph{Parallel Computing}, vol.~27, no.~14, 2001.

\bibitem{chen2010}
Y.~Chen, X.~Cui, and H.~Mei, ``Large-scale {FFT} on {GPU} clusters,'' in
  \emph{Proceedings of the 24th ACM International Conference on
  Supercomputing}, ser. ICS '10.\hskip 1em plus 0.5em minus 0.4em\relax ACM,
  2010.

\bibitem{gu2011}
L.~Gu, J.~Siegel, and X.~Li, ``Using {GPUs} to compute large out-of-card
  {FFTs},'' in \emph{Proceedings of the International Conference on
  Supercomputing}, ser. ICS '11.\hskip 1em plus 0.5em minus 0.4em\relax ACM,
  2011.

\bibitem{wu2014}
J.~Wu and J.~JaJa, ``Optimized {FFT} computations on heterogeneous platforms
  with application to the {Poisson} equation,'' \emph{Journal of Parallel and
  Distributed Computing}, vol.~74, no.~8, 2014.

\bibitem{naik2015analysis}
V.~H. Naik and C.~S. Kusur, ``Analysis of performance enhancement on graphic
  processor based heterogeneous architecture: A {CUDA} and {MATLAB}
  experiment,'' in \emph{Parallel Computing Technologies (PARCOMPTECH), 2015
  National Conference on}.\hskip 1em plus 0.5em minus 0.4em\relax IEEE, 2015,
  pp. 1--5.

\bibitem{FFTW}
\BIBentryALTinterwordspacing
FFTW, ``Fastest fourier transform in the west,'' 2018. [Online]. Available:
  \url{http://www.fftw.org/}
\BIBentrySTDinterwordspacing

\bibitem{PFFTW}
\BIBentryALTinterwordspacing
PFFTW, ``Parallel {FFTW},'' 2018. [Online]. Available:
  \url{http://www.fftw.org/fftw2_doc/fftw_4.html}
\BIBentrySTDinterwordspacing

\bibitem{pekurovsky2012p3dfft}
D.~Pekurovsky, ``{P3DFFT}: A framework for parallel computations of fourier
  transforms in three dimensions,'' \emph{SIAM Journal on Scientific
  Computing}, vol.~34, no.~4, pp. C192--C209, 2012.

\bibitem{li20102decomp}
N.~Li and S.~Laizet, ``{2DECOMP} and {FFT} - a highly scalable {2D}
  decomposition library and {FFT} interface,'' in \emph{Cray User Group 2010
  conference}, 2010, pp. 1--13.

\bibitem{Duy2014}
T.~V.~T. Duy and T.~Ozaki, ``A decomposition method with minimum communication
  amount for parallelization of multi-dimensional {FFT}s,'' \emph{Computer
  Physics Communications}, vol. 185, no.~1, pp. 153 -- 164, 2014.

\bibitem{IntelMKLFFT}
\BIBentryALTinterwordspacing
I.~Corporation, ``Intel {MKL FFT} - fast fourier transforms,'' 2018. [Online].
  Available: \url{https://software.intel.com/en-us/mkl/features/fft}
\BIBentrySTDinterwordspacing

\bibitem{Lastovetsky2007}
A.~Lastovetsky and R.~Reddy, ``Data partitioning with a functional performance
  model of heterogeneous processors,'' \emph{International Journal of High
  Performance Computing Applications}, vol.~21, no.~1, pp. 76--90, 2007.

\bibitem{ogata2008efficient}
Y.~Ogata, T.~Endo, N.~Maruyama, and S.~Matsuoka, ``An efficient, model-based
  {CPU-GPU} heterogeneous {FFT} library,'' in \emph{Parallel and Distributed
  Processing, 2008. IPDPS 2008. IEEE International Symposium on}.\hskip 1em
  plus 0.5em minus 0.4em\relax IEEE, 2008, pp. 1--10.

\bibitem{linderman2008merge}
M.~D. Linderman, J.~D. Collins, H.~Wang, and T.~H. Meng, ``Merge: a programming
  model for heterogeneous multi-core systems,'' in \emph{ACM SIGOPS operating
  systems review}, vol.~42, no.~2.\hskip 1em plus 0.5em minus 0.4em\relax ACM,
  2008, pp. 287--296.

\bibitem{quintana2009solving}
G.~Quintana-Ort{\'\i}, F.~D. Igual, E.~S. Quintana-Ort{\'\i}, and R.~A. Van~de
  Geijn, ``Solving dense linear systems on platforms with multiple hardware
  accelerators,'' in \emph{ACM Sigplan Notices}, vol.~44, no.~4.\hskip 1em plus
  0.5em minus 0.4em\relax ACM, 2009, pp. 121--130.

\bibitem{augonnet2009automatic}
C.~Augonnet, S.~Thibault, and R.~Namyst, ``Automatic calibration of performance
  models on heterogeneous multicore architectures,'' in \emph{European
  Conference on Parallel Processing}.\hskip 1em plus 0.5em minus 0.4em\relax
  Springer, 2009, pp. 56--65.

\bibitem{cybenko1989dynamic}
G.~Cybenko, ``Dynamic load balancing for distributed memory multiprocessors,''
  \emph{Journal of parallel and distributed computing}, vol.~7, no.~2, pp.
  279--301, 1989.

\bibitem{bahi2005dynamic}
J.~M. Bahi, S.~Contassot-Vivier, and R.~Couturier, ``Dynamic load balancing and
  efficient load estimators for asynchronous iterative algorithms,'' \emph{IEEE
  transactions on parallel and distributed systems}, vol.~16, no.~4, pp.
  289--299, 2005.

\bibitem{Carino2008}
R.~L. Cari{\~n}o and I.~Banicescu, ``Dynamic load balancing with adaptive
  factoring methods in scientific applications,'' \emph{The Journal of
  Supercomputing}, vol.~44, no.~1, pp. 41--63, 2008.

\bibitem{Martinez2011}
J.~A. Mart\'{\i}nez, E.~M. Garz\'{o}n, A.~Plaza, and I.~Garc\'{\i}a,
  ``Automatic tuning of iterative computation on heterogeneous multiprocessors
  with {ADITHE},'' \emph{J. Supercomput.}, vol.~58, no.~2, Nov. 2011.

\bibitem{bahi2005synchronous}
J.~Bahi, R.~Couturier, and F.~Vernier, ``Synchronous distributed load balancing
  on dynamic networks,'' \emph{Journal of Parallel and Distributed Computing},
  vol.~65, no.~11, pp. 1397--1405, 2005.

\bibitem{liu2016asynchronous}
F.~Liu, Y.~Chen, and W.~S. Wong, ``An asynchronous load balancing scheme for
  multi-server systems,'' in \emph{Ubiquitous Computing, Electronics \& Mobile
  Communication Conference (UEMCON), IEEE Annual}.\hskip 1em plus 0.5em minus
  0.4em\relax IEEE, 2016, pp. 1--7.

\bibitem{Lastovetsky2004}
A.~L. Lastovetsky and R.~Reddy, ``Data partitioning with a realistic
  performance model of networks of heterogeneous computers,'' in \emph{Parallel
  and Distributed Processing Symposium, 2004. Proceedings. 18th
  International}.\hskip 1em plus 0.5em minus 0.4em\relax IEEE, 2004, p. 104.

\bibitem{Lastovetsky2015a}
A.~L. Lastovetsky, L.~Szustak, and R.~Wyrzykowski, ``Model-based optimization
  of {MPDATA} on {Intel Xeon Phi} through load imbalancing,'' \emph{CoRR}, vol.
  abs/1507.01265, 2015.

\bibitem{Lastovetsky2015b}
A.~Lastovetsky, L.~Szustak, and R.~Wyrzykowski, ``Model-based optimization of
  {EULAG} kernel on {Intel Xeon Phi} through load imbalancing,'' \emph{IEEE
  Transactions on Parallel and Distributed Systems}, 08/2016.

\bibitem{Reddy2016}
R.~Reddy and A.~Lastovetsky, ``Bi-objective optimization of data-parallel
  applications on homogeneous multicore clusters for performance and energy,''
  \emph{IEEE Transactions on Computers}, vol.~64, no.~2, pp. 160--177, 2017.

\bibitem{hclfft}
\BIBentryALTinterwordspacing
S.~Khokhriakov and R.~Reddy, ``{HCLFFT}: Novel model-based methods for
  performance optimization of {2D} discrete {Fourier} transform on multicore
  processors,'' 2018. [Online]. Available:
  \url{https://git.ucd.ie/manumachu/hclfft.git}
\BIBentrySTDinterwordspacing

\end{thebibliography}

\appendices

\section{Transpose Routine invoked in \emph{PFFT-FPM} and \emph{PFFT-FPM-PAD}} \label{transpose}

The following routine, \emph{hcl\_tranpose\_block}, performs in-place transpose of a complex 2D square matrix of size $n \times n$. We use a block size of 64 in our experiments.

\begin{figure}
\lstset{language=C++}
\begin{lstlisting}
void hcl_transpose_scalar_block(
    fftw_complex* X1,
    fftw_complex* X2,
    const int i, const int j,
    const int n,
    const int block_size)
{
    int p, q;

    for (p = 0; p < min(n-i,block_size); p++) {
        for (q = 0; q < min(n-j,block_size); q++) {
           double tmpr = X1[p*n+q][0];
           double tmpi = X1[p*n+q][1];
           X1[p*n+q][0] = X2[q*n+p][0];
           X1[p*n+q][1] = X2[q*n+p][1];
           X2[q*n+p][0] = tmpr;
           X2[q*n+p][1] = tmpi;
        }
    }
}

void hcl_transpose_block(
    fftw_complex* X,
    const int start, const int end,
    const int n,
    const unsigned int nt,
    const int block_size)
{
    int i, j;

#pragma omp parallel for shared(X) private(i, j) num_threads(nt)
    for (i = 0; i < end; i += block_size) {
        for (j = 0; j < end; j += block_size) {
            hcl_transpose_scalar_block(
                &X[start + i*n + j],
                &X[start + j*n + i], 
                i, j, n, block_size);
        }
    }
}

\end{lstlisting}
\caption{Transpose of square matrix of size $n \times n$ using blocking.}
\end{figure}

\end{document}